\documentclass[aps,reprint,superscriptaddress,amsmath,amssymb,longbibliography]{revtex4-1}

\usepackage{graphicx}
\usepackage{amsmath}
\usepackage{bm}

\usepackage{color}
\definecolor{CLBlue}{rgb}{0, .3, .6}

\begin{document}


\title{Statistical physics of large-scale neural activity with loops}


\author{David P. Carcamo}
\affiliation{Department of Physics, Yale University, New Haven, 06511, CT, USA}
\affiliation{Quantitative Biology Institute, Yale University, New Haven, 06511, CT, USA}
\author{Christopher W. Lynn}
\affiliation{Department of Physics, Yale University, New Haven, 06511, CT, USA}
\affiliation{Quantitative Biology Institute, Yale University, New Haven, 06511, CT, USA}
\affiliation{Wu Tsai Institute, Yale University, New Haven, 06510, CT, USA}



\date{December 23, 2024}

\begin{abstract}

As experiments advance to record from tens of thousands of neurons, statistical physics provides a framework for understanding how collective activity emerges from networks of fine-scale correlations. While modeling these populations is tractable in loop-free networks, neural circuitry inherently contains feedback loops of connectivity. Here, for a class of networks with loops, we present an exact solution to the maximum entropy problem that scales to very large systems. This solution provides direct access to information-theoretic measures like the entropy of the model and the information contained in correlations, which are usually inaccessible at large scales. In turn, this allows us to search for the optimal network of correlations that contains the maximum information about population activity. Applying these methods to 45 recordings of approximately 10,000 neurons in the mouse visual system, we demonstrate that our framework captures more information---providing a better description of the population---than existing methods without loops. For a given population, our models perform even better during visual stimulation than spontaneous activity; however, the inferred interactions overlap significantly, suggesting an underlying neural circuitry that remains consistent across stimuli. Generally, we construct an optimized framework for studying the statistical physics of large neural populations, with future applications extending to other biological networks.

\end{abstract}


\maketitle

\section*{Introduction}

Statistical physics provides a powerful framework for studying how collective neural activity emerges from the vast webs of correlations between neurons \cite{wiener_nonlinear_1966, cooper_possible_1973, little_existence_1996,hopfield_neural_1982,amit_modeling_1989,hertz_introduction_1991}. Inverting these methods, one can infer the statistical interactions that explain the correlations between neurons measured in experiments \cite{schneidman_weak_2006, nguyen_inverse_2017}. This approach has provided key insights into the simple local rules underlying patterns of neural activity and information processing \cite{meshulam_collective_2017, tkacik_thermodynamics_2015, marre2009prediction, lynn_exactly_2023, lynn_exact_2023, meshulam_successes_2023, ashourvan2021pairwise, rosch2024spontaneous}. Recently, advances in two-photon microscopy and electrophysiological recordings have produced experiments capturing the simultaneous activity of thousands to tens of thousands of neurons \cite{urai2022large, gauthier2018dedicated, stringer_high-dimensional_2019, steinmetz_neuropixels_2021, demas_high-speed_2021, chung_high-density_2019, manley2024simultaneous}. As experiments grow, the number of correlations explodes exponentially. This presents a fundamental challenge: How can we identify the optimal correlations that provide the best description of a system? Solving this problem is crucial for understanding the statistical structure of neural activity at the large scales accessible in modern experiments.

Given a set of correlations, the maximum entropy principle defines the unique model that matches these correlations but contains no other sources of order \cite{jaynes1957information, thomas_m_cover_elements_2006}. This allows us to convert experimental measurements into predictive models, but it does not tell us which correlations we should include in our model to start with. Quite generally, the optimal set of correlations (that yields the best description of a system) is the network that produces the maximum entropy model with minimum entropy. This \textit{minimax entropy principle}, which remains largely unexplored, provides the framework for identifying the most important network of correlations within a system \cite{zhu_minimax_1997}. By focusing on networks without loops, many statistical physics problems---including minimax entropy---become exactly solvable, opening the door for investigations of large populations \cite{baxter2016exactly, lynn_exact_2023, lynn_exactly_2023}. However, this severely restricts the structure of correlations that we can study, and it is widely recognized that loops of connectivity between neurons play a crucial role in functional units within the brain \cite{bullmore_complex_2009, lin2023network, lynn2024heavy, bullmore_economy_2012, wang_neurophysiological_2010, lynn_physics_2019}.

Here, for a class of networks with loops, we develop a framework for identifying the most important correlations in large-scale experiments. First, by pushing exact methods to their mathematical limit, we solve the maximum entropy problem for a class of networks with loops. Second, using tools from network science, we introduce a greedy algorithm for uncovering the most important network of correlations, thus providing a locally optimal solution to the more general minimax entropy problem. We apply our framework to populations of approximately 10,000 neurons across 45 recordings of the visual system in different mice \cite{stringer_high-dimensional_2019}. In every population, we identify networks of strong correlations that capture large amounts of information about system activity. These networks produce more accurate models than loop-free networks and are consistent across different visual stimuli. Together, these results indicate that small sets of strong correlations play a critical role in guiding neural activity, and that these strong correlations are underpinned by direct neural interactions. Generally, our framework provides the tools needed to investigate the statistical structure of large populations in rapidly growing experiments.

\section*{Minimax Entropy Principle}

\begin{figure*}
\centering
    \includegraphics{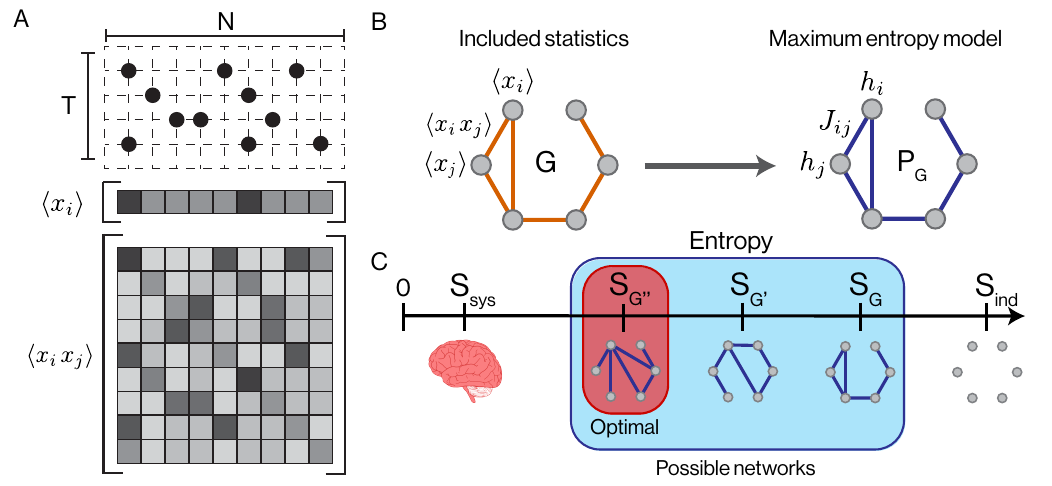}
    \caption{Identifying optimal networks of correlations. (A) Illustration of activity time series data with $N$ neurons and $T$ samples. Each point represents an active neuron within one sample. From experimental data, we can compute statistics like average activities $\langle x_i\rangle$ or pairwise correlations $\langle x_ix_j\rangle$ (see Materials and Methods). (B) A set of pairwise correlations $\langle x_ix_j\rangle$ defines a network $G$. The most unbiased model that matches these correlations and the averages $\langle x_i\rangle$ is the maximum entropy model $P_G$, which is equivalent to an Ising model with external fields $h_i$ and interactions $J_{ij}$. (C) Each set of correlations---that is, each network $G$---induces a maximum entropy $S_G$ that lies between the entropy of independent neurons $S_\text{ind}$ and the true entropy of the system $S_\text{sys}$ (to which we do not have access). Among possible networks $G$, the optimal one induces the minimum entropy $S_G$. This optimal maximum entropy model $P_G$ is the minimax entropy model.}
    \label{fig:maxEnt}
\end{figure*}

For a system of $N$ neurons $i = 1,\hdots, N$, experiments give us access to samples of collective activity $\bm{x} = \{x_i\}$ (Fig.~\ref{fig:maxEnt}A), where the state of each neuron naturally binarizes into active ($x_i = 1$) or silent ($x_i = 0$). The distribution over these states $P(\bm{x})$ contains all of the information about patterns of collective activity. But because the number of states grows exponentially with $N$, we cannot estimate $P(\bm{x})$ directly from data; instead, we can compute statistics like the average activities of the neurons $\langle x_i \rangle_\text{exp}$ or the correlations between pairs of neurons $\langle x_i x_j \rangle_\text{exp}$ (Fig.~\ref{fig:maxEnt}A; see Materials and Methods). If we focus on a subset of the pairwise correlations, we define a network $G$ with a node $i$ for each neuron and and an edge $(ij)$ for each correlation (Fig.~\ref{fig:maxEnt}B). Given this network of statistics, the most unbiased description of the system is the maximum entropy model
\begin{equation}
\label{eq_PG}
P_G(\bm{x}) = \frac{1}{Z} \exp\Big(\sum_i h_i x_i + \sum_{(ij)\in G} J_{ij} x_i x_j\Big),
\end{equation}
where $Z$ is the normalizing partition function, and the parameters $h_i$ and $J_{ij}$ must be computed so the model matches the experimental averages $\langle x_i\rangle_\text{exp}$ and correlations $\langle x_ix_j\rangle_\text{exp}$ for $(ij)\in G$ \cite{jaynes1957information, thomas_m_cover_elements_2006}. This model is mathematically equivalent to an Ising model from statistical mechanics with external fields $h_i$ and interactions $J_{ij}$ with network structure $G$ (Fig.~\ref{fig:maxEnt}B); this equivalence will become crucial as we extend Eq.~[\ref{eq_PG}] to large systems.

The maximum entropy principle provides the unique model that matches a set of experimental statistics and nothing else, but how should we select the most important statistics to begin with? Among all networks of correlations $G$, we would like to find the one that produces the most accurate description of the system. Specifically, we can choose $G$ to maximize the log-likelihood of the model $P_G$, or, equivalently, minimize the KL divergence with the data $D_\text{KL}(P_\text{exp}||P_G)$, where $P_\text{exp}$ is the experimental distribution over states. Due to the special form of $P_G$ in Eq.~[\ref{eq_PG}], this KL divergence simplifies to a difference in entropies,
\begin{equation}
    D_{\text{KL}}  \left(P_\text{exp}||P_G\right)  = S_G - S_{\text{exp}}\geq 0,
\end{equation}
where $S_G$ and $S_{\text{exp}}$ are the entropies of the model and the data, respectively (see Materials and Methods). Therefore, the optimal network $G$ (which minimizes the KL divergence) is the one that produces the maximum entropy model $P_G$ with the minimum entropy $S_G$ (Fig.~\ref{fig:maxEnt}C). This minimax entropy principle was originally proposed in the context of machine learning, but has received almost no attention in the study of biological systems \cite{zhu_minimax_1997}.

In addition to providing the most accurate description of the system, the optimal network $G$ can also be viewed as containing the maximum information about system activity. When we include a network of correlations $G$ in our model, our uncertainty about the system is reduced by an amount $I_G = S_\text{ind} - S_G \ge 0$, where $S_\text{ind}$ is the entropy of independent neurons. This is precisely the amount of information that the correlations in $G$ capture about the distribution over states. Therefore, by minimizing the entropy $S_G$, the optimal network not only minimizes the KL divergence with the data, it also maximizes the information $I_G$ \cite{lynn_exact_2023, lynn_exactly_2023}.

\section*{Exact Models with Loops}

While the minimax entropy principle determines the most informative correlations, which yield the most accurate description of a system, in practice we must overcome two distinct challenges. First, for each network $G$, we must compute the entropy of the maximum entropy model $S_G$. In general, computing the entropy exactly requires summing over all $2^N$ states of the system, limiting us to small systems of $N\lesssim 20$ neurons; and even approximating the entropy for larger systems is notoriously difficult \cite{strong1998entropy}. Second, even if we can compute the entropy $S_G$ for a given network $G$, we still need to search over all possible networks---that is, all combinations of correlations---to choose the model with the lowest entropy. This is a combinatorial optimization problem with a search space that explodes super-exponentially with the number of neurons $N$ \cite{korte2011combinatorial}.

\begin{figure}[t!]
\centering
\includegraphics[width=\linewidth]{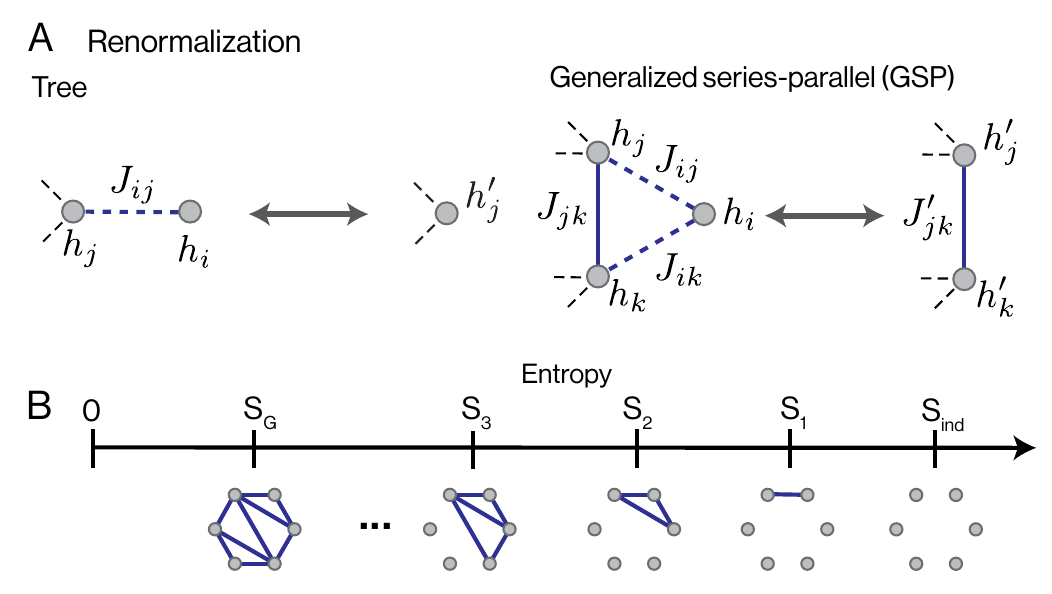}
\caption{Exact renormalization and greedy algorithm. (A) For a model $P_G$ with interactions that form a tree, one can compute the partition function $Z$ (and thus all other statistics) by iteratively summing over nodes with one connection (left). We extend exact renormalization to sum over nodes with two connections (right; see Materials and Methods). This allows us to solve the maximum entropy problem exactly for a more general class of networks known as generalized series-parallel (GSP) networks. Moreover, these techniques cannot be extended further, making GSP networks the most general class of models that can be solved with exact renormalization. (B) Greedy algorithm for constructing the minimax entropy model. Beginning with an empty network (right), we iteratively connect one new neuron to two previously-connected neurons. At each step, we choose the new connections that locally minimize the entropy of the model. Once all neurons have been connected, we arrive at our minimax entropy network $G$ (left).}
\label{fig:GSP}
\end{figure}

In statistical physics, many difficult problems become tractable if the interactions $J_{ij}$ do not contain loops \cite{baxter2016exactly}. Indeed, it was recently shown that the minimax entropy problem can be solved exactly and efficiently for networks $G$ without loops, known as trees \cite{lynn_exact_2023,lynn_exactly_2023}. The key insight is that the partition function $Z$ in Eq.~[\ref{eq_PG}] can be computed by iteratively summing over neurons $i$ with one connection (Fig.~\ref{fig:GSP}A). This process, known as exact renormalization \cite{rosten2012fundamentals}, is thought to be possible only if the network does not contain loops, as in one-dimensional Ising models or on Bethe lattices \cite{baxter2016exactly, nguyen_inverse_2017}.

Here we extend exact renormalization to a more general class of networks with loops (see Materials and Methods). In particular, rather than summing over neurons with only one connection, one can compute $Z$ by iteratively summing over neurons with two connections (Fig.~\ref{fig:GSP}A). This is possible for a class of networks $G$ known as generalized series-parallel (GSP) networks, which include trees, planar graphs, and series-parallel networks \cite{duffin1965topology}. Moreover, this procedure cannot be extended further, making GSP networks the most general class of models that can be solved with exact renormalization (see Materials and Methods). Once $Z$ is calculated, one can then compute all of the statistics in the model by taking derivatives of the form $\langle x_i\rangle = \frac{\partial \log Z}{\partial h_i}$ and $\langle x_ix_j\rangle = \frac{\partial \log Z}{\partial J_{ij}}$. Inverting this procedure, one can begin with experimental averages $\langle x_i\rangle_\text{exp}$ and correlations $\langle x_ix_j\rangle_\text{exp}$ on a GSP network $G$ and compute the corresponding model parameters $h_i$ and $J_{ij}$ (see Materials and Methods). This solves the maximum entropy problem both exactly (that is, without approximations) and efficiently, paving the way for applications to very large systems.

\section*{Greedy Algorithm}

Using the above techniques, we can construct exact maximum entropy models $P_G$ for a class of networks with loops. But we still need to search over all possible GSP networks $G$ to find the one that provides the best description of the system, minimizing the entropy $S_G$. Searching by brute force is impossible for all but the smallest of networks, so we instead decompose the search into a sequence of local optimization problems \cite{jungnickel2005graphs}. This decomposition is made possible by the fact that any GSP network can be constructed by repeatedly connecting one new neuron $i$ to two previously connected neurons $j$ and $k$. At each step of this growth process, connecting neuron $i$ to neurons $j$ and $k$ means that we are adding the correlations $\langle x_ix_j\rangle_\text{exp}$ and $\langle x_ix_k\rangle_\text{exp}$ to the constraints in our model. Including these correlations decreases the entropy of the model by an amount
\begin{equation}
\Delta S_i = S(x_i) + S(x_j,x_k) - S_\text{pair}(x_i,x_j,x_k),
\label{eq_dS}
\end{equation}
where $S(\cdot)$ represents the experimental entropy, and $S_\text{pair}(\cdot)$ represents the entropy in our model (that is, the maximum entropy consistent with the means and pairwise correlations between $i$, $j$, and $k$; see Materials and Methods). Thus, for any GSP network $G$, we are able to exactly compute the entropy of the model $P_G$ by combining each of these contributions,
\begin{equation}
S_G = S_\text{ind} - \sum_i \Delta S_i.
\label{eq_SG}
\end{equation}

We are now prepared to write down a greedy algorithm that constructs a locally optimal GSP network by minimizing the entropy at each step (Fig.~\ref{fig:GSP}B):
\begin{enumerate}
    \item Beginning with a model of independent neurons (defined by a network with no connections), we connect the pair of neurons that results in the lowest entropy; this is the pair of neurons with the largest mutual information \cite{lynn_exact_2023, lynn_exactly_2023}.
    \item We then iteratively connect a new neuron $i$ to two previously connected neurons $j$ and $k$ so as to maximize the entropy drop $\Delta S_i$ in Eq.~[\ref{eq_dS}].
    \item This process continues until all neurons have been added to the network.
\end{enumerate}
Using this greedy algorithm, we construct a GSP network $G$ that approximately minimizes the entropy $S_G$, thus capturing as much information $I_G$ about the system as possible. Testing on simulated networks of up to $N=10,000$ neurons, this algorithm correctly identifies over $75\%$ of the ground-truth interactions $J_{ij}$ and captures over $98\%$ of the total information $I_G$ in the population (see Supporting Information). Together, our renormalization procedure and greedy algorithm combine to produce an exact maximum entropy model $P_G$ that is optimized to provide the best description of an experimental system.

\section*{Modeling Large-scale Neural Activity}

The efficiency of our minimax entropy framework gives us the opportunity to study populations of neurons at the vast scales accessible in modern experiments. Each GSP network, however, only contains $2N-3$ correlations, while the total number of pairwise correlations grows quadratically as $N(N-1)/2$. This means that as $N$ grows in large experiments, we can only include a vanishingly small fraction $\sim$$4/N$ of all the pairwise correlations in any model; and even if we fit all of these, there is still no guarantee that we can predict higher-order correlations between three or more neurons. Can such a sparse network of correlations have any hope of capturing a macroscocpic fraction of the information in the data?

\begin{figure}
    \centering
    \includegraphics[width=\linewidth]{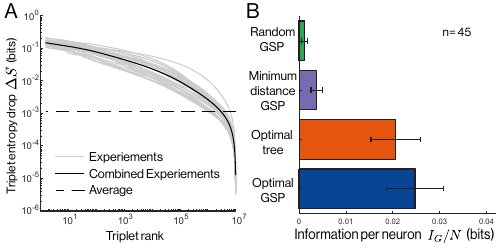}
    \caption{Information captured by different networks. (A) Distributions of entropy drops $\Delta S$ (Eq.~[\ref{eq_dS}]) for triplets of neurons. Grey lines represent the 45 different recordings, the black line defines the average distribution across recordings, and the dashed line illustrates the average entropy drop $\overline{\Delta S}$ across all recordings and neuron triplets. (B) Information per neuron captured by different networks of correlations. Values and error bars represent averages and standard deviations across the 45 different recordings.}
    \label{fig:Accuracy}
\end{figure}

To answer this question, we apply our framework to 45 recordings of $N\approx 10,000$ neurons in the mouse visual system, taken from seven different mice in previous experiments \cite{stringer_high-dimensional_2019}. The activity of each neuron is measured using two-photon calcium imaging at a rate of approximately $1.5$ Hz and binarized to reflect activity significantly above baseline (see Materials and Methods). For such large experiments, each GSP network only includes $\sim$$4/N\approx 0.04\%$ of all the correlations between pairs of neurons. For such a sparse network to have any predictive power, we need a small number of correlations to contain an unusually large amount of information $I_G$. From Eqs.~[\ref{eq_dS}-\ref{eq_SG}], we see that the information contained in any GSP network $G$ can be decomposed into a sum of contributions from neuron triplets, $I_G = S_\text{ind} - S_G = \sum_i \Delta S_i$. Averaging across all experiments, we find a mean entropy drop of $\overline{\Delta S} = 0.001$ bits; this defines the amount of information (per neuron) contained in a typical GSP network, $I_G/N \approx \overline{\Delta S}$. However, when we look at the full distribution of entropy drops $\Delta S$, we see that it is heavy-tailed (Fig.~\ref{fig:Accuracy}A), with some rare values that are orders of magnitude larger than average. This indicates that our minimax entropy framework may be particularly will suited for this neural data.

Across the 45 different experiments, we find that random GSP networks only capture $I_G/N = 0.001$ bits per neuron, as predicted by the approximation $I_G/N \approx \overline{\Delta S}$ (Fig.~\ref{fig:Accuracy}B). By contrast, our greedy algorithm identifies locally optimal networks that contain $I_G/N = 0.025$ bits per neuron, over twenty times more information than a typical network. These optimized models reduce our total uncertainty about each neuron by $I_G/S_\text{ind} = 10.9\%$ (compared to $I_G/S_\text{ind} = 0.5\%$ for random networks), a remarkable amount considering that each network only includes two correlations per neuron. Thus, across multiple recordings, we consistently identify sparse backbones of correlations that capture large amounts of information about the neural activity.

To better understand these important correlations, we can compare against other types of networks. For example, one might suspect that neurons interact most strongly with their nearest neighbors. To test this hypothesis, we can construct GSP networks that connect the physically closest neurons in each recording; these ``minimum distance" correlations only capture $I_G/N = 0.004$ bits per neuron (Fig.~\ref{fig:Accuracy}B). Similarly, building upon past results, we can construct the most informative trees of correlations \cite{lynn_exact_2023,lynn_exactly_2023}. Across all recordings, these optimal trees capture less information than our locally optimal GSP networks (Fig.~\ref{fig:Accuracy}B). Together, these results indicate that the most important correlations include long-range connections and loops of connectivity.

\begin{figure*}

\centering
    \includegraphics[width=17.8cm]{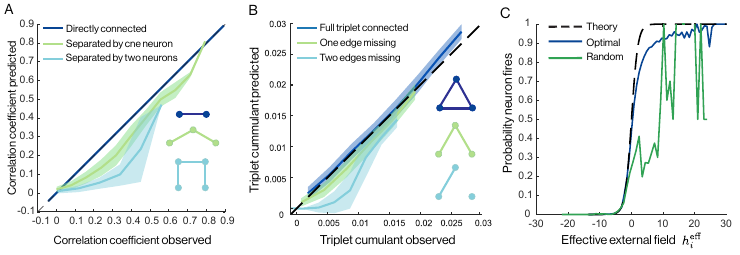}
    \caption{Predicting features of neural activity. (A) For one recording of $N = 10,144$ neurons, we compare the correlation coefficients between pairs of neurons measured in data versus predicted by the minimax entropy model $P_G$. Different colors represent pairs of neurons separated by different distances in the network $G$, and the dashed line indicates equality. Plots are computed by binning neuron pairs along the x-axis, with lines and shaded regions representing means and standard deviations within each bin. (B) Cumulants among triplets of neurons measured in data versus predicted by our minimax entropy model. Different colors represent triplets of neurons with different numbers of pairwise correlations constrained in the model. Lines and shaded regions represent means and standard deviations within bins along the x-axis. (C) Probability of a neuron $i$ being active as a function of the effective field $h_i^\text{eff}(\bm{x})$ computed in the minimax entropy model (blue) and for a random network (green). Plots are averaged over all neurons within the population, and the dashed line represents the analytic prediction in Eq.~[\ref{eq_Pspike}].}
    \label{fig:Predictions}
\end{figure*}

\section*{Predicting Correlations in Neural Activity}

By maximizing the information $I_G$, we hope to arrive at a model $P_G$ that can be used to predict structure in the neural activity. In general, making exact predictions in the Ising model (Eq.~[\ref{eq_PG}]) is infeasible, and approximations require time-consuming Monte Carlo simulations. Here, by generalizing the famous Bethe solution for trees \cite{baxter2016exactly}, we derive exact and efficient model predictions for our class of GSP networks (see Materials and Methods).

In Fig.~\ref{fig:Predictions}A, we compare the experimental correlation coefficients between neurons with the values predicted by our minimax entropy model $P_G$. We focus on a single population of $N = 10,144$ neurons recorded while the mouse is exposed to a sequence of natural images \cite{stringer_high-dimensional_2019}. For pairs of neurons $i$ and $j$ connected in the network $G$, the model exactly matches the observed correlations, as desired. For neurons separated by one intermediate neuron, the model still provides reasonable predictions, even for very strong correlations. This demonstrates that some pairwise correlations can be explained as arising indirectly through shared correlations with a third neuron. As we increase the distance between neurons in the network, the model predictions become less accurate, indicating that indirect interactions via two or more intermediate neurons are not sufficient to explain the observed correlations.

In addition to pairwise statistics, we can also investigate higher-order correlations among groups of neurons. In Fig.~\ref{fig:Predictions}B, we compare the triplet cumulants $\langle(x_i - \langle x_i\rangle)(x_j - \langle x_j\rangle)(x_k - \langle x_k\rangle)\rangle$ predicted by our model $P_G$ with those measured in experiment. For triplets that are fully connected in $G$---such that all three pairwise correlations are constrained in $P_G$---the model accurately predicts the triplet correlations within experimental errors. Even with one correlation missing, our model still accurately predicts the triplet cumulants. By contrast, a random set of pairwise correlations (that is, a typical network $G$) provides almost no predictive power about these higher-order correlations (see Supporting Information).

At the level of individual cells, each model $P_G$ makes a clear prediction for the response of neuron $i$ to the state of the rest of the population,
\begin{equation}
P_G(x_i = 1\,|\, \bm{x}) = \big( 1 + \exp(-h_i^\text{eff}(\bm{x}))\big)^{-1},
\label{eq_Pspike}
\end{equation}
where $h_i^\text{eff}(\bm{x}) = h_i + \sum_{j\in G_i}J_{ij}x_j$ is the effective field on neuron $i$ and $G_i$ denotes the neighbors of $i$ in $G$ \cite{lynn_exact_2023, lynn_exactly_2023, meshulam_successes_2023}. In Fig.~\ref{fig:Predictions}C, we see that a random set of correlations is insufficient to predict the responses of individual neurons. Meanwhile, our optimized GSP network exhibits good agreement with data across a wide range of spike probabilities. Together, the results of Fig.~\ref{fig:Predictions} demonstrate that our minimax entropy framework, despite only including two correlations per neuron, is capable of predicting key features of collective neural activity. This is only possible because we select the most informative correlations in the population.

\begin{figure*}

\centering
    \includegraphics[width=17.8cm]{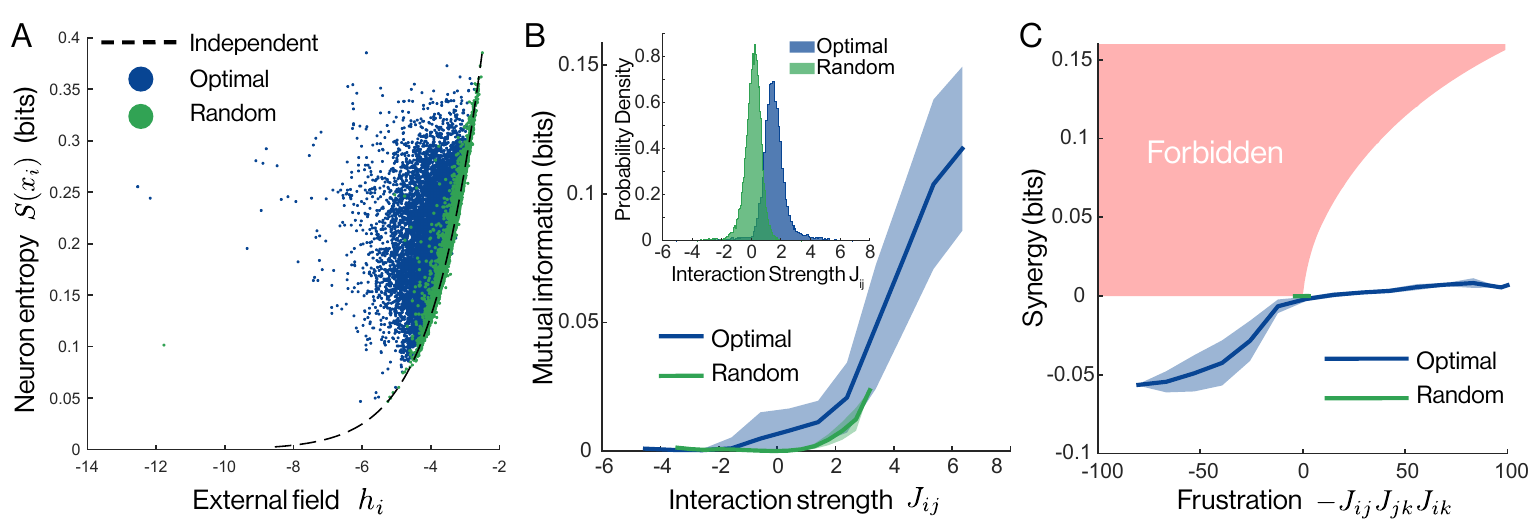}
    \caption{Structure of optimal model. (A) Neuron entropies $S(x_i)$ versus external fields $h_i$ computed in the optimal model (blue) and a random network (green). Dashed line illustrates the independent prediction. (B) Mutual information between neurons versus the inferred interaction $J_{ij}$ for different models. (B, Inset) Distribution of inferred interactions $J_{ij}$ for the optimal model and a random network. (C) Synergy versus frustration $-J_{ij}J_{jk}J_{ki}$ among triplets of neurons for different models. The red region cannot be attained by any Ising triplet.}
    \label{fig:Model}
\end{figure*}

\section*{Structure of Minimax Entropy Model}

Given a GSP network of correlations $G$, we provide the tools to exactly infer the maximum model $P_G$ in Eq.~[\ref{eq_PG}] (see Materials and Methods), which is equivalent to an Ising model with binary states $x_i \in \{0,1\}$. Leveraging this connection to statistical physics, we can investigate the different types of models produced by different networks $G$. For example, each external field $h_i$ represents the individual bias for neuron $i$ towards activity or silence. For an independent neuron, this field precisely defines the average activity $\langle x_i\rangle = 1/(1 + e^{-h_i})$ and therefore the independent entropy $S(x_i)$. For a random network $G$, because the correlations are so weak, the entropy of each neuron closely follows this independent prediction (Fig.~\ref{fig:Model}A). Meanwhile, the minimax entropy framework identifies a strong network of correlations, leading to neuron entropies $S(x_i)$ that significantly differ from independence.

In each model, the interactions $J_{ij}$ represent the influence of neuron $i$ to induce activity ($J_{ij} > 0$) or silence ($J_{ij} < 0$) in neuron $j$, and vice versa. For a random network, these interactions are evenly split between positive and negative (Fig.~\ref{fig:Model}B, Inset), yielding a description that is akin to the Sherrington-Kirkpatrick model of a spin glass \cite{sherrington1975solvable}. The most informative correlations, by contrast, produce almost exclusively positive interactions. This makes the minimax entropy model $P_G$ similar to an Ising ferromagnet, in which positive interactions can build upon one another to generate large-scale order and long-range correlations \cite{newell1953theory, brush1967history}.

Using our exact solution to the maximum entropy problem, we can gain insight into how the interactions $J_{ij}$ relate to the information $I_G$ that a network captures about neural activity. For a GSP network $G$, we derive the following decomposition of the information into non-negative components (see Materials and Methods),
\begin{equation}
I_G = \sum_{(ij)\in G} I(x_i,x_j) + \sum_{(ijk)\in G} \text{Syn}(x_i,x_j,x_k),
\label{eq_IG}
\end{equation}
where $I(x_i,x_j) = S(x_i) + S(x_j) - S(x_i,x_j)$ is the mutual information between neurons $i$ and $j$, and the second sum runs over all triplets $(ijk)$ that form a triangle in $G$. Inside the final sum is the synergy $\text{Syn}(x_i,x_j,x_k)$ (see Materials and Methods), which represents the amount of information that two neurons contain about a third above and beyond their mutual information \cite{schneidman_network_2003,schneidman_synergy_2003}. This decomposition tells us that optimal GSP network $G$ should focus on pairs of neurons with large mutual informations and triplets with large synergies.

The interactions $J_{ij}$ in our optimized network, in addition to being mostly positive, are also much stronger than those in the random network. In Fig.~\ref{fig:Model}B, we see that these strongly positive interactions produce pairs of neurons with large mutual informations, as desired. For comparison, synergy increases if the interactions between neurons present competing influences \cite{schneidman_network_2003,schneidman_synergy_2003}; in the Ising model, competing interactions give rise to frustration, which we can quantify using the product $-J_{ij}J_{jk}J_{ki}$. We note that synergy and frustration can only arise in networks with loops, and therefore cannot be studied using previous methods on trees \cite{lynn_exact_2023, lynn_exactly_2023}. In GSP networks, we find that positive synergy can only be achieved by frustrated triplets (Fig.~\ref{fig:Model}C). However, in our minimax entropy model, we find that most triplets have negative synergy, such that neurons contain redundant information about one another (Fig.~\ref{fig:Model}C). These results demonstrate that to maximize the information $I_G$ in this neural population, the optimal model focuses on pairs of neurons with large mutual informations (underpinned by strongly positive interactions), even at the expense of synergistic information.

\section*{Effects of Visual Stimulation}

In the visual cortex, neural activity is strongly driven by details in the visual world \cite{stringer_high-dimensional_2019, gilbert2013top}. Yet these patterns of activity are also shaped by recurrent connections, which form feedback loops of interactions that do not change from one stimulus to another \cite{ko2013emergence, hofer2011differential, smith2008spatial}. This raises a clear question: Do the most important correlations between neurons remain consistent across stimuli, or do they depend crucially on the visual scene?

\begin{figure}[t!]

\centering
\includegraphics[width=\linewidth]{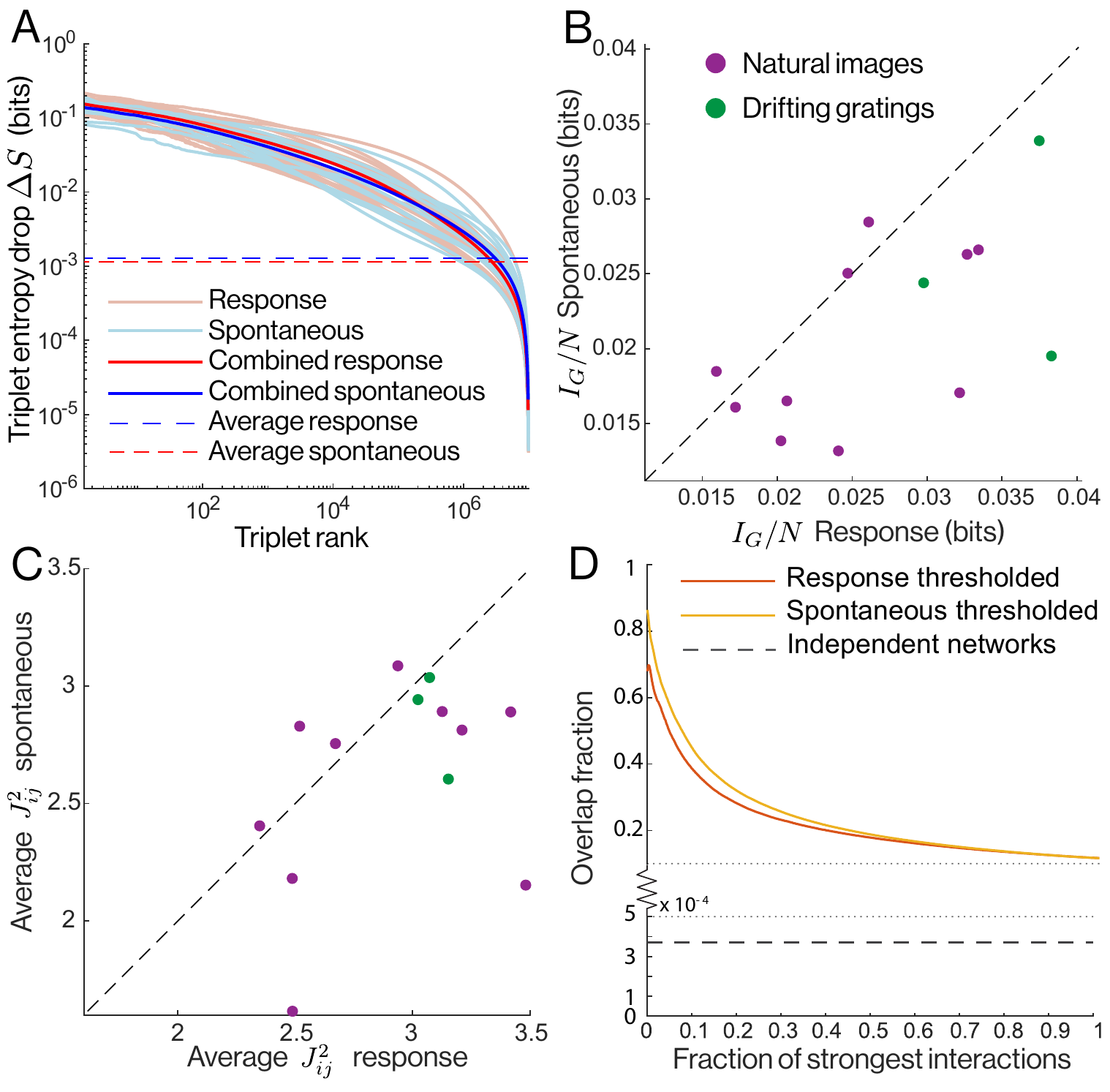}
\caption{Effects of visual stimulation on optimal correlations. (A) Distributions of entropy drops $\Delta S$ (Eq.~[\ref{eq_dS}]) for triplets of neurons during visual stimulation (red) and spontaneous activity (blue). Dark lines define the average distributions for each stimulus condition, and the dashed lines illustrate the average entropy drops $\overline{\Delta S}$. (B) Information per neuron $I_G/N$ captured by the optimal networks $G$ for visual responses and spontaneous activity. Each point represents a unique neural population, colors indicate the type of visual stimulus, and the dashed line defines equality. (C) Average squared interaction strengths $J_{ij}^2$ in the minimax entropy models $P_G$ fit to the same populations during visual responses or spontaneous activity. (D) Fractional overlap between optimal networks during visual responses and spontaneous activity. For each population, we either threshold the visual response network (red) or the spontaneous network (yellow) based on the strongest inferred interactions $J_{ij}$ and leave the other network fixed to compute the overlap. Two independent networks would have an overlap of $\sim$$4/N$, which for $N\approx 10,000$ is $0.04\%$ (dashed line).}
\label{fig:Comparison}
\end{figure}

Out of the 45 different recordings of $N\approx 10,000$ neurons (Fig.~\ref{fig:Accuracy}), 26 correspond to populations that were recorded twice---once in response to visual stimuli (either natural images or drifting gratings) and once during spontaneous activity (with a grey or black screen). Across triplets of neurons, we find almost identical distributions of entropy drops $\Delta S$ (Eq.~[\ref{eq_dS}]) between responding and spontaneous activity (Fig.~\ref{fig:Comparison}A). This suggests that for any GSP network $G$, the amount of information contained in the correlations $I_G = \sum_i \Delta S_i$ will be consistent across the two conditions. Indeed, for random networks, we find that the typical information per neuron $I_G/N\approx \overline{\Delta S}$ does not vary significantly between conditions (see Supporting Information). However, focusing on the networks identified by our minimax entropy framework, the most important correlations in the populations contain $30\%$ more information when responding to visual stimuli than in spontaneous activity (Fig.~\ref{fig:Comparison}B). This means that, for the same neurons and the same number of correlations, one can achieve a better description of the neural activity when the population is driven by visual cues.

To understand this difference in information, we can study the structures of the optimal models $P_G$. For responses to visual stimuli, we find that the inferred interactions between neurons $J_{ij}$ are stronger than for spontaneous activity (Fig.~\ref{fig:Comparison}C). This increase in interaction strength corresponds to an increase in the mutual informations between neurons, as seen previously (Fig.~\ref{fig:Model}B). Yet despite these differences in interaction strengths, the optimal networks themselves remain remarkably consistent between conditions. For two independent networks with $N = 10,000$ nodes and $2N$ connections, we expect a fractional overlap of $\sim$$4/N = 0.04\%$. By contrast, the most informative correlations within each population exhibit an overlap of $12\%$ between responding and spontaneous activity, over two orders of magnitude more than independent networks (Fig.~\ref{fig:Comparison}D). This overlap grows even larger as we focus on stronger interactions $J_{ij}$ within the optimal networks (Fig.~\ref{fig:Comparison}D), reaching $70-85\%$ for the strongest interactions \cite{hoshal_stimulus_2024}.

Together, these results demonstrate that (i) the information contained in correlations increases when responding to visual stimuli (Fig.~\ref{fig:Comparison}B); but (ii) the networks formed by these correlations remain strikingly consistent from spontaneous to stimulated activity (Fig.~\ref{fig:Comparison}D). In turn, this suggests that the most important correlations in the visual system may driven by actual interactions between neurons---thus remaining consistent across stimuli---but that these correlations are amplified when responding to visual cues.

\section*{Discussion}


As experimental techniques advance, enabling simultaneous recordings of larger and larger populations of neurons \cite{urai2022large, gauthier2018dedicated, stringer_high-dimensional_2019, steinmetz_neuropixels_2021, demas_high-speed_2021, chung_high-density_2019, manley2024simultaneous}, we face new challenges in extracting meaningful statistical structure at vast scales. The primary difficulty lies in constructing quantitative models that can be used to predict the probabilities of high-dimensional patterns of activity. While recent techniques from statistical physics have solved this problem in models without loops of connectivity \cite{lynn_exact_2023, lynn_exactly_2023}, the cortex is known to exhibit complex circuits of recurrent connections between neurons \cite{bullmore_complex_2009, lin2023network, lynn2024heavy, bullmore_economy_2012, wang_neurophysiological_2010, lynn_physics_2019}.

Here, for a class of models with loops, we present an exact solution to the maximum entropy problem that scales to very large systems. This solution gives us direct access to information-theoretic quantities like the entropy of the model and the amount of information that it captures about the system, which are usually inaccessible at large scales (Fig.~\ref{fig:maxEnt}). In turn, this allows us to search for the model that provides the best description of the data, and we present a locally optimal algorithm for executing this search (Fig.~\ref{fig:GSP}). The end result is a framework for (i) identifying the most important correlations within large neuronal populations and (ii) using these correlations to make exact predictions about collective activity.

We apply our methods to 45 recordings of approximately $10,000$ neurons in the mouse visual cortex \cite{stringer_high-dimensional_2019}. In each recording, we identify optimal correlations that contain over twenty times more information than typical networks (Fig.~\ref{fig:Accuracy}). This information allows us to quantitatively predict additional correlations between pairs and triplets of neurons that were not included in the model (Fig.~\ref{fig:Predictions}). Notably, the optimal correlations in a population capture more information during visual stimulation than spontaneous activity; however, the networks formed by these correlations remain strikingly consistent, hinting at a common underlying neural circuitry (Fig.~\ref{fig:Comparison}).

Broadly, we present a framework---based on the little-known minimax entropy principle \cite{baxter2016exactly, lynn_exact_2023, lynn_exactly_2023}---for constructing optimized statistical models of the large populations becoming accessible in modern experiments. These methods are general, applying to any system with binary data. This opens the door for future investigations into collective neural activity in other systems, species, and imaging modalities \cite{lynn_physics_2019, tkacik_thermodynamics_2015, marre2009prediction, meshulam_successes_2023, lynn_exact_2023, lynn_exactly_2023, ashourvan2021pairwise, rosch2024spontaneous, urai2022large, gauthier2018dedicated, stringer_high-dimensional_2019, steinmetz_neuropixels_2021, demas_high-speed_2021, chung_high-density_2019, manley2024simultaneous}. One can also use the same techniques to study collective behaviors in other complex living systems, such as genetic interactions, chromatin structure, and animal behaviors \cite{lynn_surges_2019, Shi-01, Messelink-01, Lezon-01, Dixit-01, Weigt-01, Marks-01, Bialek-01, Bialek-02, Mora-01}. Finally, our exact maximum entropy solution provides the foundation for the future development of improved approximate models, for example based on mean-field techniques and cluster expansions \cite{yedidia2005constructing, tanaka2000information, cocco_adaptive_2011, cocco_adaptive_2012}. In this way, our minimax entropy framework provides a principled starting point for statistical models (with loops) of large-scale neural activity.

\appendix

\section*{Materials and Methods}

\subsection*{Data}

The data consists of calcium imaging recordings of populations of $N= 10506 \pm 1737$ (mean $\pm$ standard deviation) from the mouse visual system at a sampling rate of about 1.5 Hz, measured in previous experiments \cite{stringer_high-dimensional_2019}. We study 45 recordings of 7 separate mice who were free to run on an air-floating ball as images were presented on three computer screens. Stimuli included natural images, distorted natural images, drifting gratings, and grey screens (to measure spontaneous activity). These stimuli were presented to the mice on average $T= 4570 \pm 1486$ times during each recording, and the sampling of neural activity across all neurons aligns with the stimulus presentations. The recordings can be broadly divided into two groups: responses to stimuli and spontaneous activity. There are 13 instances of neuronal populations being recorded both during spontaneous activity and while responding to stimuli (10 for natural images and 3 for drifting gratings). These pairs of recordings capture the same sets of neurons in the same mice under different stimulus conditions.

The activity of each neuron was binarized into active ($x_i = 1$) or silent ($x_i = 0$) at each moment of time based on whether or not its calcium trace reached two standard deviations above its mean activity. The collective activity is then defined by the binary vector $\bm{x} = \{x_i\}$. Due to the large number of neurons, there are pairs of neurons that never fire together during a recording. When computing experimental averages, we correct for this by adding one pseudo-count, such that the experimental  statistics are given by
\begin{align}
    \left< x_i \right> &= \frac{1}{1+T}\left(1+\sum_{t=1}^Tx_i(t)\right),\\
    \left< x_i x_j\right> &=\frac{1}{1+T}\left(1+\sum_{t=1}^Tx_i(t)x_j(t)\right),
\end{align}
where $x_i(t)$ defines the activity of neuron $i$ at time $t$.

\subsection*{Maximum Entropy Principle}

The maximum entropy principle determines the least biased model that matches a specified set of statistical constraints \cite{jaynes1957information, thomas_m_cover_elements_2006}. Here, we focus on a model constrained to match the empirical averages of neural activity $\langle x_i\rangle$ and a subset of the pairwise correlations $\langle x_ix_j\rangle$ that lie on a network $G$. The maximum entropy model consistent with these constraints takes the form of an Ising model with interactions that lie on the network $G$ (Eq.~[\ref{eq_PG}]). For an all-to-all network $G$, one arrives at the pairwise maximum entropy model, which has provided key insights into the collective behavior of smaller populations of up to $N\sim 100$ neurons \cite{schneidman_weak_2006, nguyen_inverse_2017, meshulam_collective_2017, tkacik_thermodynamics_2015, meshulam_successes_2023}.

\subsection*{Partition Function}

For an Ising model with interactions that lie on a GSP network $G$ [Eq.~(\ref{eq_PG})], we provide an exact solution for the statistics $\langle x_i\rangle$ and $\langle x_ix_j\rangle$ as functions of the parameters $h_i$ and $J_{ij}$. As a first step, we compute the partition function $Z$. Introducing a zero-point energy $f = 0$, which will soon become useful, the Boltzmann distribution takes the form
\begin{equation}
\label{eq_PT}
P_G(\bm{x}) = \frac{1}{Z}\text{exp}\Bigg[\sum_{(ij)\in G} J_{ij}x_ix_j + \sum_i h_ix_i + f\Bigg].
\end{equation}
To compute the partition function,
\begin{equation}
\label{eq_Z}
Z = \sum_{\bm{x}} \text{exp}\Bigg[\sum_{(ij)\in G} J_{ij}x_ix_j + \sum_i h_ix_i + f\Bigg],
\end{equation}
we start by summing over one variable. Our goal is to find a new system of $N-1$ variables with the same partition function $Z$. If we can repeat this process until no variables remain, then computing $Z$ will be trivial.

We label the nodes $i = 1,\hdots, N$ based on the order that they are removed (or summed over), and we let, $J^{(i)}_{jk}$, $h^{(i)}_i$, and $f^{(i)}$ denote the updated parameters at step $i$. Consider summing over a variable $i$ with only two connections in the network, say to variables $j$ and $k$, which themselves are connected (such that $J_{jk} \neq 0$). In GSP networks, such a node $i$ is always guaranteed to exist. To keep the partition function fixed, the new system with $i$ removed must satisfy the equations
\begin{multline}
\label{eq_sum}
e^{J_{jk}^{(i)}x_jx_k+h_j^{(i)}x_j+h_k^{(i)}x_k + f^{(i)}}\big(e^{J_{ij}^{(i)}x_j + J_{ik}^{(i)}x_k + h^{(i)}_i }+ 1\big) = \\e^{J_{jk}^{(i+1)}x_jx_k+h_j^{(i+1)}x_j+h_k^{(i+1)}x_k + f^{(i+1)}}.
\end{multline}
This is a system of four equations (one for each value of $x_j$ and $x_k$), which we can solve for the new parameters
\begin{align}
    f^{(i+1)} &= f^{(i)} + \ln \left( e^{h_i^{(i)}} + 1\right),\label{EQ:consitency1}\\
    h_j^{(i+1)} &= h_j^{(i)} - \ln \left( e^{h_i^{(i)}} + 1\right) + \ln                \left( e^{J_{ij}^{(i)} + h_i^{(i)}} + 1\right) , \label{EQ:consitency2}\\
    h_k^{(i+1)} &= h_k^{(i)} - \ln \left( e^{h_i^{(i)}} + 1\right) + \ln                \left( e^{J_{ik}^{(i)} + h_i^{(i)}} + 1\right) ,\label{EQ:consitency3} \\
    J_{jk}^{(i+1)} &= J_{jk}^{(i)} + \ln \left( e^{h_i^{(i)}} + 1\right) - \ln                \left( e^{J_{ij}^{(i)} + h_i^{(i)}} + 1\right) \label{EQ:consitency4} \\ & - \ln  \left( e^{J_{ik}^{(i)} + h_i^{(i)}} + 1\right) + \ln \left( e^{J_{ij}^{(i)} + J_{ik}^{(i)} + h_i^{(i)}} + 1\right).\notag
\end{align}
After repeating the above procedure $N$ times, we have summed over all nodes, and we are left with a single parameter $\mathcal{F} = f^{(N+1)}$. This is the negative free energy, and the partition function is given by
\begin{equation}
Z = e^{\mathcal{F}}.
\end{equation}

At each step, we have assumed that we can find a node $i$ with only two connections to nodes $j$ and $k$ that are themselves connected. After removing $i$, we must find another such node to repeat the calculation. The class of networks for which this process can continue down to a final root node are precisely the set of GSP networks \cite{korneyenko_combinatorial_1994}. Moreover, we note that this is the furthest we can push this technique. Attempting to remove any node with three neighbors would lead to an overdetermined system of 8 equations and 7 parameters. In this case, one would need to introduce an additional triplet interaction between the tree neighbors, and we would diverge from the realm of Ising models. We therefore establish that GSP networks are the most general class of networks that can be solved through exact renormalization.

\subsection*{Average Activities and Correlations}

To compute statistics, we take derivatives of the partition function,
\begin{align}
\langle x_i\rangle &= \frac{d \ln Z}{d h_i} = \frac{d\mathcal{F}}{d h_i}, \\
\langle x_ix_j\rangle &= \frac{d \ln Z}{d J_{ij}} = \frac{d\mathcal{F}}{d J_{ij}},
\end{align}
where $\frac{d}{d h_i}$ and $\frac{d}{d J_{ij}}$ represent total derivatives, which account for indirect dependencies via Eqs.~[\ref{EQ:consitency1}]-[\ref{EQ:consitency4}]. Since $\frac{d\mathcal{F}}{d f^{(i+1)}} = 1$ and $\frac{d h_i^{(i)}}{d h_i} = 1$, the above procedure yields
\begin{multline}
\langle x_i\rangle =\frac{\partial f^{(i+1)}}{\partial h^{(i)}_i} + \frac{d\mathcal{F}}{d h^{(i+1)}_j} \frac{\partial h^{(i+1)}_j}{\partial h^{(i)}_i} + \frac{d\mathcal{F}}{d h^{(i+1)}_k}  \frac{\partial h^{(i+1)}_k}{\partial h^{(i)}_i} \\+ \frac{d\mathcal{F}}{d J^{(i+1)}_{jk}}  \frac{\partial J^{(i+1)}_{jk}}{\partial h^{(i)}_i}.
\end{multline}
Noticing that
\begin{align}
 \frac{d \mathcal{F}}{d h^{(i+1)}_j} &=  \frac{d \mathcal{F}}{d h_j} = \langle x_j\rangle,\\
 \frac{d \mathcal{F}}{d h^{(i+1)}_k} &=  \frac{d \mathcal{F}}{d h_k} = \langle x_k\rangle,  \\
 \frac{d \mathcal{F}}{d J^{(i+1)}_{jk}} &=  \frac{d \mathcal{F}}{d J_{jk}} = \langle x_j x_k\rangle,
\end{align}
and taking derivatives of Eqs.~[\ref{EQ:consitency1}]-[\ref{EQ:consitency4}], we have

\begin{multline}
    \left< x_i \right> = \frac{1 - \left< x_j \right> - \left< x_k \right> + \left< x_j x_k \right> }{1+e^{- h^{(i)}_i}} + \frac{\left< x_j \right> -  \left< x_j x_k \right> }{1+e^{- J^{(i)}_{ij}- h^{(i)}_i }} \\ +\frac{\left< x_k \right> -  \left< x_j x_k \right> }{1+e^{- J^{(i)}_{ik}- h^{(i)}_i }} + \frac{\left< x_j x_k \right> }{1+e^{- J^{(i)}_{ij}- J^{(i)}_{ik}- h^{(i)}_i }}.
    \label{EQ:meanGSP}
\end{multline}
The correlations follow analogously,
\begin{equation}
    \left< x_i x_j \right> = \frac{\left< x_j \right> -  \left< x_j x_k \right> }{1+e^{- J^{(i)}_{ij}- h^{(i)}_i }} + \frac{\left< x_j x_k \right> }{1+e^{- J^{(i)}_{ij}- J^{(i)}_{ik}- h^{(i)}_i }},
    \label{EQ:corrij}
\end{equation}
\begin{equation}
    \left< x_i x_k \right> = \frac{\left< x_k \right> -  \left< x_j x_k \right> }{1+e^{- J^{(i)}_{ik}- h^{(i)}_i }} + \frac{\left< x_j x_k \right> }{1+e^{- J^{(i)}_{ij}- J^{(i)}_{ik}- h^{(i)}_i }}.
    \label{EQ:corrik}
\end{equation}
Thus, by iterating through the nodes in the opposite order from which they were summed over to compute $Z$, we can compute the average activities $\langle x_i\rangle$ and correlations $\langle x_ix_j\rangle$ for pairs $(ij)\in G$. For the correlations $\langle x_ix_j\rangle$ that are not in the network (that is, for $(ij) \not\in G$), see Supporting Information.

\subsection*{Maximum Entropy Solution}

We have solved the ``forward" problem for an Ising model on a GSP network. Now we seek to solve the ``inverse" (or maximum entropy) problem for the parameters $h_i$ and $J_{ij}$ as functions of the observed statistics $\langle x_i\rangle$ and $\langle x_ix_j\rangle$. In practice, this amounts to inverting Eqs.~[\ref{EQ:meanGSP}]-[\ref{EQ:corrik}]. We start with the last node in the decimation order $i = N$ and calculate its external field from its empirical average as 
\begin{equation}
    h_N^{(N)} =\ln\left( \frac{\langle x_N \rangle}{1- \langle x_N \rangle} \right).
\end{equation}
Next, we connect node $i = N - 1$ to node $i = N$, yielding the parameters
\begin{equation}
    h_{N-1}^{(N-1)} = \ln\left( \frac{\langle x_N\rangle - \langle x_{N,N-1}\rangle}{1+\langle x_{N,N-1}\rangle- \langle x_N\rangle - \langle x_{N-1}\rangle} \right),
\end{equation}
\begin{equation}
     J_{N,N-1}^{(N-1)} \ln\left( \frac{\langle x_{N,N-1}\rangle}{\langle x_N\rangle - \langle x_{N,N-1}\rangle}\right) - h_{N-1}^{(N-1)}.
\end{equation}
We must also update the external field on node $i = N$,
\begin{multline}
    h_N^{(N-1)} = h_N^{(N)} + \ln\left(e^{h_{N-1}^{(N-1)}}+1 \right) \\- \ln\left(e^{h_{N-1}^{(N-1)}+J_{N,N-1}^{(N-1)}}+1 \right).
\end{multline}
For the remaining nodes $i = N-2,\hdots, 1$, we must invert Eqs.~[\ref{EQ:meanGSP}]-[\ref{EQ:corrik}] numerically to calculate $h_i^{(i)}$, $J_{ij}^{(i)}$,$J_{ik}^{(i)}$ in terms of $\langle x_i \rangle$, $\langle x_i x_j\rangle$, and $\langle x_i, x_k\rangle$, where $j$ and $k $ are the parents of $i$. We can then use Eqs.~[\ref{EQ:consitency1}]-[\ref{EQ:consitency4}] to update $h^{(i)}_j$, $h^{(i)}_k$,  $J^{(i)}_{jk}$. This process continues until all nodes have been added to the network, and we arrive at the solution $h_i = h_i^{(1)}$ and $J_{ij} = J_{ij}^{(1)}$.

\subsection*{Minimax Entropy}

For a given network $G$, the difference between the maximum entropy distribution $P_G$ and the experimental distribution $P_\text{exp}$ is quantified by the Kullback–Leibler (KL) divergence,
\begin{align}
&D_\text{KL}(P_\text{exp}||P_G) = \left\langle \ln \left( \frac{P_\text{exp}(\bm{x})}{P_G(\bm{x})} \right) \right\rangle_\text{exp} \\
&=  - S_{\text{exp}}+ \ln(Z) - \sum_{(ij)\in G} J_{ij} \left< x_i x_j \right>_{\text{exp}} - \sum_i h_i \left<x_i \right>_{\text{exp}}\\
&= - S_{\text{exp}}+\ln(Z)  -\sum_{(ij) \in G} J_{ij} \left< x_i x_j \right> - \sum_i  h_i \left<x_i \right> \\
&= S_G - S_{\text{exp}},
\end{align}
where the penultimate equality follows from the maximum entropy constraints $\langle x_i\rangle = \langle x_i\rangle_\text{exp}$ and $\langle x_ix_j\rangle = \langle x_ix_j\rangle_\text{exp}$. The equation above tells us that the optimal network $G$, which minimizes the KL divergence from the data, is the one that minimizes the entropy $S_G$ of the maximum entropy model. This is the minimax entropy principle, which was discovered over 20 years ago \cite{zhu_minimax_1997}, but remains largely unexplored in the study of complex living systems \cite{lynn_exactly_2023, lynn_exact_2023}.

\subsection*{Greedy Algorithm}

Directly searching over all possible GSP networks is computationally intractable. Instead, we will take a \textit{greedy} approach to minimizing entropy. As discussed above, the class of GSP networks is the set of networks that you can grow by iteratively adding a new node and connecting it to two existing nodes that are already connected (Fig.~\ref{fig:GSP}B). This definition leads directly to a greedy algorithm for constructing the optimal network: At each step of the network construction, we should connect a new node $i$ to two existing nodes $j$ and $k$ (that are already connected) so as to minimize the entropy $S_G$.

To implement this greedy algorithm, we need to compute the drop in entropy from connecting a new node $i$ to two existing nodes $j$ and $k$; this is precisely the drop in entropy from fitting the correlations $\langle x_ix_j\rangle$ and $\langle x_ix_k\rangle$ in the maximum entropy model. Before connecting $i$ in the network, the distribution over states factorizes,
\begin{equation}
P_G(\bm{x}) = P_G(x_i) P_G(\bm{x}_{-i}),
\end{equation}
where $\bm{x}_{-i}$ denotes the states of all variables other than $i$. Since $P_G$ matches the average $\langle x_i\rangle$, we note that $P_G(x_i)$ is the same as the experimental marginal $P(x_i)$. After connecting $i$ to $j$ and $k$, we arrive at a new network $G'$ with a distribution of the form
\begin{equation}
P_{G'}(\bm{x}) = P_{G'}(x_i|x_j,x_k) P_{G'}(\bm{x}_{-i}).
\end{equation}
From our decimation procedure above, we know that $P_{G'}(\bm{x}_{-i}) = \sum_{x_i}P_{G'}(\bm{x}) = P_G(\bm{x}_{-i})$. The drop in entropy thus reduces to
\begin{align}
\Delta S_i &= S_G - S_{G'} \\
&= S(x_i) + S_G(\bm{x}_{-i}) - S_{G'}(x_i|x_j,x_k) - S_G(\bm{x}_{-i}) \\
\label{EQ:dS}
&= S(x_i) + S(x_i,x_j) - S_\text{pair}(x_i,x_j,x_k),
\end{align}
where $S(\cdot)$ denotes the experimental entropy and $S_\text{pair}(\cdot)$ represents the entropy of the maximum entropy model that is consistent with the averages and all pairwise correlations between variables.

We have arrived at our greedy algorithm. At each step, we consider all combinations of new nodes $i$ and pairs $j$ and $k$ that are already connected in the network. For each triplet, we compute the entropy drop in Eq.~[\ref{EQ:dS}], and for the largest drop, we connect $i$ to $j$ and $k$. We then repeat this process until all nodes have been connected in the network. By minimizing the entropy $S_G$ at each step, this greedy algorithm provides a locally optimal solution to the minimax entropy problem (Fig.~\ref{fig:GSP}).

\subsection*{Decomposing Information}

Finally, we derive a decomposition of the information contained within a GSP network of correlations $G$. As discussed above, the information contained in any network of correlations is the drop in entropy $I_G = S_\text{ind} -S_G$, where $S_G$ is the entropy of the maximum entropy model $P_G$. For a GSP network $G$, we showed in the previous section that this information can be decomposed into a sequence of entropy drops
\begin{equation}
\label{EQ:Infodrop}
I_G = \sum_i \Delta S_i = \sum_i \Big( S(x_i) + S(x_i,x_j) - S_\text{pair}(x_i,x_j,x_k) \Big),
\end{equation}
where $j$ and $k$ are the parents of $i$ in the network construction.

The decomposition in Eq.~[\ref{EQ:Infodrop}] depends on the order in which we add nodes to the network during construction. We will now derive a new decomposition that does not depend on this choice of order. To begin, we introduce a new quantity known as the \textit{synergy},
\begin{multline}
\text{Syn}(x_i,x_j,x_k) = S(x_i)+S(x_j)+S(x_k) - S_\text{pair}(x_i,x_j,x_k)\\
-I(x_i,x_j)-I(x_i,x_k)-I(x_j,x_k).
\label{EQ:Synergy}
\end{multline}
The synergy represents the amount of information that two variables contain about a third beyond their pairwise dependencies \cite{brenner_synergy_2000,schneidman_synergy_2003}. We note that the above synergy is computed in the model $P_G$, not the experimental distribution $P_\text{exp}$. We also note that the synergy is symmetric under permutations of $x_i$, $x_j$, and $x_k$. Substituting into Eq.~[\ref{EQ:Infodrop}], we have
\begin{equation}
        I_G = \sum_i \text{Syn}(x_i,x_j,x_k) + I(x_i,x_j) + I(x_i,x_k). 
\end{equation}

When a new variable $i$ is added to the network, we create two new edges (connecting $i$ to $j$ and $k$) and one new triangle (among $i$, $j$, and $k$). We therefore see that the above sum can be rewritten as a sum over network motifs: edges and triangles. This gives us a decomposition of the information that does not depend on our choice of node order in the network construction,
\begin{equation}
I_G = \sum_{(ij) \in G} I(x_i,x_j) + \sum_{(ijk)\in G} \text{Syn}(x_i,x_j,x_k),
\end{equation}
where the first sum runs over all edges in $G$, and the second sum runs over all fully-connected triplets of neurons (triangles) in $G$. This decomposition for GSP networks generalizes previous decompositions of the information contained in trees of correlations, or networks without loops \cite{lynn_exact_2023,lynn_exactly_2023}.


\bibliography{ExactPNAS_cwl1}

\end{document}




\SItext

\subsection*{Computing All Correlations}

The goal of this section is to calculate the averages $\langle x_i\rangle$ and pairwise correlations $\langle x_i x_j\rangle$  of binary variables drawn from an Ising Model,  $P_G(x) = \frac{1}{Z} \exp \left( \sum_{(i,j) \in G} J_{ij} x_i x_j + \sum_i h_i x_i \right) $, with external fields $h_i$ and interactions $J_{ij}$ that lie on a generalized series-parallel (GSP) network $G$. In the main text, we show how to calculate averages $\langle x_i\rangle$ and pairwise correlations $\langle x_i x_j \rangle$ where $J_{ij} \neq 0$. We now want to calculate the full correlation matrix, $\left< x_i x_j \right> $ for all remaining pairwise correlations. We begin by iteratively decimating each node to compute the partition function Z, as described in the main text. We can label each node by the order in which they were decimated, $i = 1,...,N$. Here we show how to compute the correlations $\langle x_i x_j \rangle$ not in the network; that is for $(ij \notin G)$. To begin, we assume that we have computed the correlations $\langle x_i x_k \rangle$ for all nodes $k>i>j$. Then, if we compute $\langle x_i x_j\rangle$ the procedure will follow by induction. 

We can directly calculate the covariance between these pair of nodes as
\begin{equation}
    \frac{d \left< x_i\right>}{dh_j} = \left< x_i x_j\right> - \left< x_i\right> \left< x_j\right>.
\end{equation}
From our main text, we calculate $\langle x_i\rangle$ as
\begin{equation}
    \left< x_i \right> = \frac{1 - \left< x_{p_1(i)} \right> - \left< x_{p_2(i)} \right> + \left< x_{p_1(i)} x_{p_2(i)} \right> }{1+e^{- h^{(i)}_i}} + \frac{\left< x_{p_1(i)} \right> -  \left< x_{p_1(i)} x_{p_2(i)} \right> }{1+e^{- J^{(i)}_{i{p_1(i)}}- h^{(i)}_i }} \\ +\frac{\left< x_{p_2(i)} \right> -  \left< x_{p_1(i)} x_{p_2(i)} \right> }{1+e^{- J^{(i)}_{i{p_2(i)}}- h^{(i)}_i }} + \frac{\left< x_{p_1(i)}x_{p_2(i)} \right> }{1+e^{- J^{(i)}_{i{p_1(i)}}- J^{(i)}_{i{p_2(i)}}- h^{(i)}_i }}.
    \label{EQ:meanGSP}
\end{equation}
We will use ${p_1(i)}$ and $p_2(i)$ to be the parent neurons of neuron $j$ (the two neighboring nodes when neuron $i$ was removed during the calculation of $Z$). We can calculate the derivative of $\left< x_i \right>$ with respect to $h_j$ by applying the chain rule to the equation for $\left< x_i \right>$, leading to
\begin{multline}
     \frac{d \left< x_i\right>}{dh_j} = \frac{\partial \left< x_i\right>}{\partial h_i^{(i)}}\frac{d h_i^{(i)}}{d h_j}+\frac{\partial\left< x_i\right>}{\partial J_{ip_1(i)}^{(i)}}\frac{d J_{ip_1(i)}^{(i)}}{d h_j}+\frac{\partial\left< x_i\right>}{\partial J_{ip_2(i)}^{(i)}}\frac{d J_{ip_2(i)}^{(i)}}{d h_j}+\frac{\partial \left< x_i\right>}{\partial \left< x_{p_1(i)}\right>}\frac{d \left< x_{p_1(i)}\right>}{d h_j} \\ +\frac{\partial\left< x_i\right>}{\partial \left< x_{p_2(i)}\right>}\frac{d \left< x_{p_2(i)}\right>}{d h_j}+\frac{\partial\left< x_i\right>}{\partial \left< x_{p_1(i)} x_{p_2(i)}\right>}\frac{d\left< x_{p_1(i)} x_{p_2(i)}\right>}{d h_j}.
\end{multline}
From our main text, we can calculate the derivatives $\frac{\partial \left< x_i\right>}{\partial h_i^{(i)}},\frac{\partial\left< x_i\right>}{\partial J_{ip_1(i)}^{(i)}},\frac{\partial\left< x_i\right>}{\partial J_{ip_2(i)}^{(i)}},\frac{\partial \left< x_i\right>}{\partial \left< x_{p_1(i)}\right>},\frac{\partial\left< x_i\right>}{\partial \left< x_{p_2(i)}\right>},$ and $ \frac{\partial\left< x_i\right>}{\partial \left< x_{p_1(i)} x_{p_2(i)}\right>}$ directly from Eqs.~[12 , 20]. These derivatives are
\begin{multline}
     \frac{\partial \left< x_i\right>}{\partial h_i} = \frac{e^{-h_i^{(i)}}}{\left( 1+ e^{-h_i^{(i)}}\right)^2}\left(1-\left< x_{p_1(i)}\right>-\left< x_{p_2(i)}\right>+\left< x_{p_1(i)}x_{p_2(i)}\right>\right) \\
     + \frac{e^{-h_i^{(i)}-J_{ip_1(i)}^{(i)}}}{\left( 1+ e^{-h_i^{(i)}-J_{ip_1(i)}}\right)^2}\left(\left< x_{p_1(i)}\right>-\left< x_{p_1(i)}x_{p_2(i)}\right>\right)
     + \frac{e^{-h_i^{(i)}-J_{ip_2(i)}^{(i)}}}{\left( 1+ e^{-h_i^{(i)}-J_{ip_2(i)}^{(i)}}\right)^2}\left(\left< x_{p_2(i)}\right>-\left< x_{p_1(i)}x_{p_2(i)}\right>\right)\\
     +\frac{e^{-h_i^{(i)}-J_{ip_1(i)}^{(i)}-J_{ip_2(i)}^{(i)}}}{\left( 1+ e^{-h_i^{(i)}-J_{ip_1(i)}^{(i)}-J_{ip_2(i)}^{(i)}}\right)^2}\left< x_{p_1(i)}x_{p_2(i)}\right>,
\end{multline}
\begin{equation}
     \frac{\partial\left< x_i\right>}{\partial J_{ip_1(i)}} = \frac{e^{-h_i^{(i)}-J_{ip_1(i)}^{(i)}}}{\left( 1+ e^{-h_i^{(i)}-J_{ip_1(i)}}\right)^2}\left(\left< x_{p_1(i)}\right>-\left< x_{p_1(i)}x_{p_2(i)}\right>\right) + \frac{e^{-h_i^{(i)}-J_{ip_1(i)}^{(i)}-J_{ip_2(i)}^{(i)}}}{\left( 1+ e^{-h_i^{(i)}-J_{ip_1(i)}^{(i)}-J_{ip_2(i)}^{(i)}}\right)^2}\left< x_{p_1(i)}x_{p_2(i)}\right>,
\end{equation}
\begin{equation}
     \frac{\partial\left< x_i\right>}{\partial J_{ip_2(i)}}= \frac{e^{-h_i^{(i)}-J_{ip_2(i)}^{(i)}}}{\left( 1+ e^{-h_i^{(i)}-J_{ip_2(i)}^{(i)}}\right)^2}\left(\left< x_{p_2(i)}\right>-\left< x_{p_1(i)}x_{p_2(i)}\right>\right) +\frac{e^{-h_i^{(i)}-J_{ip_1(i)}^{(i)}-J_{ip_2(i)}^{(i)}}}{\left( 1+ e^{-h_i^{(i)}-J_{ip_1(i)}^{(i)}-J_{ip_2(i)}^{(i)}}\right)^2}\left< x_{p_1(i)}x_{p_2(i)}\right>,
\end{equation}
\begin{equation}
     \frac{\partial \left< x_i\right>}{\partial \left< x_{p_1(i)}\right>} = \frac{-1}{ 1+ e^{-h_i^{(i)}}} + \frac{1}{ 1+ e^{-h_i^{(i)}-J_{ip_1(i)}^{(i)}}},
\end{equation}
\begin{equation}
     \frac{\partial\left< x_i\right>}{\partial \left< x_{p_2(i)}\right>} = \frac{-1}{ 1+ e^{-h_i^{(i)}}} + \frac{1}{1+ e^{-h_i^{(i)}-J_{ip_2(i)}^{(i)}}} ,
\end{equation}
\begin{equation}
 \frac{\partial\left< x_i\right>}{\partial \left< x_{p_1(i)} x_{p_2(i)}\right>} = \frac{1}{ 1+ e^{-h_i^{(i)}}} - \frac{1}{ 1+ e^{-h_i^{(i)}-J_{ip_1(i)}^{(i)}}} -\frac{1}{ 1+ e^{-h_i^{(i)}-J_{ip_2(i)}^{(i)}}} +\frac{1}{ 1+ e^{-h_i^{(i)}-J_{ip_1(i)}^{(i)}-J_{ip_2(i)}^{(i)}}}.
\end{equation}
We can calculate the full derivative as
\begin{align}
     \frac{d \left< x_i\right>}{dh_j} &= \frac{e^{-h_i^{(i)}}}{\left( 1+ e^{-h_i^{(i)}}\right)^2}\left(1-\left< x_{p_1(i)}\right>-\left< x_{p_2(i)}\right>+\left< x_{p_1(i)}x_{p_2(i)}\right>\right)\frac{d h_i^{(i)}}{d h_j} \nonumber \\ 
     & + \frac{e^{-h_i^{(i)}-J_{ip_1(i)}^{(i)}}}{\left( 1+ e^{-h_i^{(i)}-J_{ip_1(i)}}\right)^2}\left(\left< x_{p_1(i)}\right>-\left< x_{p_1(i)}x_{p_2(i)}\right>\right) \left(\frac{d h_i^{(i)}}{d h_j} + \frac{d J_{ip_1(i)}^{(i)}}{d h_j} \right) \nonumber\\
     &+ \frac{e^{-h_i^{(i)}-J_{ip_2(i)}^{(i)}}}{\left( 1+ e^{-h_i^{(i)}-J_{ip_2(i)}^{(i)}}\right)^2}\left(\left< x_{p_2(i)}\right>-\left< x_{p_1(i)}x_{p_2(i)}\right>\right)\left(\frac{d h_i^{(i)}}{d h_j} + \frac{d J_{ip_2(i)}^{(i)}}{d h_j} \right) \nonumber\\ 
     &+\frac{e^{-h_i^{(i)}-J_{ip_1(i)}^{(i)}-J_{ip_2(i)}^{(i)}}}{\left( 1+ e^{-h_i^{(i)}-J_{ip_1(i)}^{(i)}-J_{ip_2(i)}^{(i)}}\right)^2}\left< x_{p_1(i)}x_{p_2(i)}\right>\left(\frac{d h_i^{(i)}}{d h_j} + \frac{d J_{ip_1(i)}^{(i)}}{d h_j} + \frac{d J_{ip_2(i)}^{(i)}}{d h_j} \right) \nonumber \\ 
     &+  \left[\frac{-1}{ 1+ e^{-h_i^{(i)}}} + \frac{1}{ 1+ e^{-h_i^{(i)}-J_{ip_1(i)}^{(i)}}} \right] \frac{d \left< x_{p_1(i)}\right>}{d h_j} \nonumber\\
     &+\left[\frac{-1}{ 1+ e^{-h_i^{(i)}}} + \frac{1}{1+ e^{-h_i^{(i)}-J_{ip_2(i)}^{(i)}}} \right]\frac{d \left< x_{p_2(i)}\right>}{d h_j} \nonumber\\
     &+\left[\frac{1}{ 1+ e^{-h_i^{(i)}}} - \frac{1}{ 1+ e^{-h_i^{(i)}-J_{ip_1(i)}^{(i)}}} -\frac{1}{ 1+ e^{-h_i^{(i)}-J_{ip_2(i)}^{(i)}}} +\frac{1}{ 1+ e^{-h_i^{(i)}-J_{ip_1(i)}^{(i)}-J_{ip_2(i)}^{(i)}}} \right]\frac{d\left< x_{p_1(i)} x_{p_2(i)}\right>}{d h_j}.
\end{align}
This leaves calculating the derivatives of $h_i^{(i)}, J_{i,p_1(i)}^{(i)}, J_{i,p_2(i)}^{(i)}, \left< x_{p_1(i)} \right>,\left< x_{p_2(i)}\right>$, and $\left< x_{p_1(i)} x_{p_2(i)}\right>$ with respect to $h_j$. We can focus on $h_i^{(i)}, J_{ip_1(i)}^{(i)}, J_{ip_2(i)}^{(i)}$ and notice that the dependence of these parameters on $h_j$ runs through $h_{p_1(j)}^{(j+1)}, h_{p_2(j)}^{(j+1)} $ and $J_{p_1(j)p_2(j)}^{(j+1)}$. Therefore, we have:

\begin{equation}
    \frac{d h_i^{(i)}}{d h_j} = \frac{d h_i^{(i)}}{d h_{p_1(j)}^{(j+1)}} \frac{d h_{p_1(j)}^{(j+1)}}{d h_j}+ \frac{d h_i^{(i)}}{d h_{p_2(j)}^{(j+1)}}\frac{d h_{p_2(j)}^{(j+1)}}{d h_j} + \frac{d h_i^{(i)}}{d J_{p_1(j)p_2(j)}^{(j+1)}}\frac{d J_{p_1(j)p_2(j)}^{(j+1)}}{d h_j},
    \label{EQ:hihj}
\end{equation}
\begin{equation}
    \frac{d J_{i p_1(i)}^{(i)}}{d h_j} = \frac{d J_{i p_1(i)}^{(i)}}{d h_{p_1(j)}^{(j+1)}} \frac{d h_{p_1(j)}^{(j+1)}}{d h_j}+ \frac{d J_{i p_1(i)}^{(i)}}{d h_{p_2(j)}^{(j+1)}}\frac{d h_{p_2(j)}^{(j+1)}}{d h_j} + \frac{d J_{i p_1(i)}^{(i)}}{d J_{p_1(j)p_2(j)}^{(j+1)}}\frac{d J_{p_1(j)p_2(j)}^{(j+1)}}{d h_j},
    \label{EQ:Jip1ihj}
\end{equation}
\begin{equation}
    \frac{d J_{i p_2(i)}^{(i)}}{d h_j} = \frac{d J_{i p_2(i)}^{(i)}}{d h_{p_1(j)}^{(j+1)}} \frac{d h_{p_1(j)}^{(j+1)}}{d h_j}+ \frac{d J_{i p_2(i)}^{(i)}}{d h_{p_2(j)}^{(j+1)}}\frac{d h_{p_2(j)}^{(j+1)}}{d h_j} + \frac{d J_{i p_2(i)}^{(i)}}{d J_{p_1(j)p_2(j)}^{(j+1)}}\frac{d J_{p_1(j)p_2(j)}^{(j+1)}}{d h_j}.
    \label{EQ:Jip2ihj}
\end{equation}
We can calculate the derivatives $\frac{d h_{p_1(j)}^{(j+1)}}{d h_j}, \frac{d h_{p_2(j)}^{(j+1)}}{d h_j}, \text{and } \frac{d J_{p_1(j) p_2(j) }^{(j+1)}}{d h_j}$,  from Eqs.~[12] in the main text.  We will eventually need the derivatives of $h_i^{(i)}, J_{i p_1(i)}^{(i)}, J_{ ip_2(i)}^{(i)}$ with respect to $J_{j p_1(j)}^{(j)}, J_{j p_2(j)}^{(j)}$:
\begin{equation}
    \frac{d h_i^{(i)}}{d J_{j p_1(j)}^{(j)}} = \frac{d h_i^{(i)}}{d h_{p_1(j)}^{(j+1)}} \frac{d h_{p_1(j)}^{(j+1)}}{d J_{j p_1(j)}^{(j)}}+ \frac{d h_i^{(i)}}{d h_{p_2(j)}^{(j+1)}}\frac{d h_{p_2(j)}^{(j+1)}}{d J_{j p_1(j)}^{(j)}} + \frac{d h_i^{(i)}}{d J_{p_1(j)p_2(j)}^{(j+1)}}\frac{d J_{p_1(j)p_2(j)}^{(j+1)}}{d J_{j p_1(j)}^{(j)}},
\end{equation}
\begin{equation}
    \frac{d J_{ip_1(i)}^{(i)}}{d J_{j p_1(j)}^{(j)}} = \frac{d J_{ip_1(i)}^{(i)}}{d h_{p_1(j)}^{(j+1)}} \frac{d h_{p_1(j)}^{(j+1)}}{d J_{j p_1(j)}^{(j)}}+ \frac{d J_{ip_1(i)}^{(i)}}{d h_{p_2(j)}^{(j+1)}}\frac{d h_{p_2(j)}^{(j+1)}}{d J_{j p_1(j)}^{(j)}} + \frac{d J_{ip_1(i)}^{(i)}}{d J_{p_1(j) p_2(j)}^{(j+1)}}\frac{d J_{p_1(j) p_2(j)}^{(j+1)}}{d J_{j p_1(j)}^{(j)}},
\end{equation}
\begin{equation}
    \frac{d J_{ip_2(i)}^{(i)}}{d J_{j p_1(j)}^{(j)}} = \frac{d J_{ip_2(i)}^{(i)}}{d h_{p_1(j)}^{(j+1)}} \frac{d h_{p_1(j)}^{(j+1)}}{d J_{j p_1(j)}^{(j)}}+ \frac{d J_{ip_2(i)}^{(i)}}{d h_{p_2(j)}^{(j+1)}}\frac{d h_{p_2(j)}^{(j+1)}}{d J_{j p_1(j)}^{(j)}} + \frac{d J_{ip_2(i)}^{(i)}}{d J_{p_1(j)p_2(j)}^{(j+1)}}\frac{d J_{p_1(j)p_2(j)}^{(j+1)}}{d J_{j p_1(j)}^{(j)}},
\end{equation}
\begin{equation}
    \frac{d h_i^{(i)}}{d J_{j p_2(j)}^{(j)}} = \frac{d h_i^{(i)}}{d h_{p_1(j)}^{(j+1)}} \frac{d h_{p_1(j)}^{(j+1)}}{d J_{j p_2(j)}^{(j)} }+ \frac{d h_i^{(i)}}{d h_{p_2(j)}^{(j+1)}}\frac{d h_{p_2(j)}^{(j+1)}}{d J_{j p_2(j)}^{(j)}} + \frac{d h_i^{(i)}}{d J_{p_1(j)p_2(j)}^{(j+1)}}\frac{d J_{p_1(j)p_2(j)}^{(j+1)}}{d J_{j p_2(j)}^{(j)}},
\end{equation}
\begin{equation}
    \frac{d J_{ip_1(i)}^{(i)}}{d J_{j p_2(j)}^{(j)}} = \frac{d J_{ip_1(i)}^{(i)}}{d h_{p_1(j)}^{(j+1)}} \frac{d h_{p_1(j)}^{(j+1)}}{d J_{j p_2(j)}^{(j)}}+ \frac{d J_{ip_1(i)}^{(i)}}{d h_{p_2(j)}^{(j+1)}}\frac{d h_{p_2(j)}^{(j+1)}}{d J_{j p_2(j)}^{(j)}} + \frac{d J_{ip_1(i)}^{(i)} }{d J_{p_1(j)p_2(j)}^{(j+1)}}\frac{d J_{p_1(j)p_2(j)}^{(j+1)}}{d J_{j p_2(j)}^{(j)}},
\end{equation}
\begin{equation}
    \frac{d J_{ip_2(i)}^{(i)}}{d J_{j p_2(j)}^{(j)}} = \frac{d J_{ip_2(i)}^{(i)}}{d h_{p_1(j)}^{(j+1)}} \frac{d h_{p_1(j)}^{(j+1)}}{d J_{j p_2(j)}^{(j)}}+ \frac{d J_{ip_2(i)}^{(i)}}{d h_{p_2(j)}^{(j+1)}}\frac{d h_{p_2(j)}^{(j+1)}}{d J_{j p_2(j)}^{(j)}} + \frac{d J_{ip_2(i)}^{(i)}}{d J_{p_1(j)p_2(j)}^{(j+1)}}\frac{d J_{p_1(j)p_2(j)}^{(j+1)}}{d J_{j p_2(j)}^{(j)}}.
\end{equation}
Since $p_1(j)>j$ and $p_2(j)>j$, we can assume we know all derivatives of $h_{i}^{(i)}, J_{ip_1(1)}^{(i)}, J_{i p_2(i)}^{(i)}$ with respect to $h_{p_1(j)}^{(j+1)}, h_{p21(j)}^{(j+1)},J_{p_1(j)p_2(j)}^{(j+1)}$. Then, to calculate Eqs.~[\ref{EQ:hihj} - \ref{EQ:Jip2ihj}], we need to compute the derivatives of $h_{p_1(j)}^{(j+1)}, h_{p_2(j)}^{(j+1)}, J_{ p_1(j) p_2(j)}^{(j+1)}$ with respect to $h_j^{(j)}, J_{j p_1(j)}^{(j)}, J_{j p_2(j)}^{(j)}$ which can be directly calculated using Eq.~[12] in the main text. These derivatives are
\begin{align*}
    \frac{d h_{p_1(j)}^{(j+1)}}{d h_j^{(j)}} &= \frac{1}{ 1+ e^{-h_j^{(j)}}} + \frac{1}{ 1+ e^{-h_j^{(j)}-J_{jp_1(j)}^{(j)}}}, \\ 
    \frac{d h_{p_1(j)}^{(j+1)}}{d J_{j p_1(j)}^{(j)}} &= \frac{1}{ 1+ e^{-h_j^{(j)}-J_{jp_1(j)}^{(j)}}},\\ 
    \frac{d h_{p_1(j)}^{(j+1)}}{d J_{j p_2(j)}^{(j)}} &= 0,\\ 
    \frac{d h_{p_2(j)}^{(j+1)}}{d h_j^{(j)}} &= \frac{1}{ 1+ e^{-h_j^{(j)}}} + \frac{1}{ 1+ e^{-h_j^{(j)}-J_{jp_2(j)}^{(j)}}},\\ 
    \frac{d h_{p_2(j)}^{(j+1)}}{d J_{j p_1(j)}^{(j)}} &= 0,\\ 
    \frac{d h_{p_2(j)}^{(j+1)}}{d J_{j p_2(j)}^{(j)}} &= \frac{1}{ 1+ e^{-h_j^{(j)}-J_{jp_2(j)}^{(j)}}}, \\ 
    \frac{d J_{ p_1(j) p_2(j)}^{(j+1)}}{d h_j^{(j)}} &= \frac{1}{ 1+ e^{-h_j^{(j)}}} - \frac{1}{ 1+ e^{-h_j^{(j)}-J_{jp_1(j)}^{(j)}}} - \frac{1}{ 1+ e^{-h_j^{(j)}-J_{jp_2(j)}^{(j)}}} + \frac{1}{ 1+ e^{-h_j^{(j)}-J_{jp_1(j)}^{(j)}-J_{jp_2(j)}^{(j)}}},\\ 
    \frac{d J_{ p_1(j) p_2(j)}^{(j+1)}}{d J_{j p_1(j)}^{(j)}} &= \frac{-1}{ 1+ e^{-h_j^{(j)}-J_{jp_1(j)}^{(j)}}} +\frac{1}{ 1+ e^{-h_j^{(j)}-J_{jp_1(j)}^{(j)}-J_{jp_2(j)}^{(j)}}},\\ 
    \frac{d J_{ p_1(j) p_2(j)}^{(j+1)}}{d J_{j p_2(j)}^{(j)}} &= \frac{-1}{ 1+ e^{-h_j^{(j)}-J_{jp_2(j)}^{(j)}}} + \frac{1}{ 1+ e^{-h_j^{(j)}-J_{jp_1(j)}^{(j)}-J_{jp_2(j)}^{(j)}}}.
\end{align*}
This completes the calculation of the derivatives of  $h_i^{(i)}, J_{ip_1(i)}^{(i)}, J_{ip_2(i)}^{(i)}$ with respect to $h_j$.

Finally, we must compute the derivatives of  $\left< x_{p_1(i)} \right>,\left< x_{p_2(i)}\right>,\left< x_{p_1(i)} x_{p_2(i)}\right>$ with respect to $h_j$. Since $p_1(i)>j$ and $p_2(i)>j$, we will assume we have already calculated the derivatives $\left< x_{p_1(i)}\right> , \left< x_{p_2(i)}\right>, \left< x_{p_1(i)} x_{p_2(i)}\right>$ with respect to $h_j$. Induction will follow if we can calculate the derivatives of $\left< x_i\right> , \left< x_i x_{p_1(i)}\right>, \left< x_i x_{p_2(i)}\right>$ with respect to $h_j$. Again applying the chain rule these derivatives become
\begin{align}
    \frac{d\left< x_i x_{p_1(i)}\right>}{d h_j} &= \frac{d\left< x_i x_{p_1(i)}\right>}{d h_i^{(i)}} \frac{d h_i^{(i)}}{ d h_j }+ \frac{d\left< x_i x_{p_1(i)}\right>}{d J_{i p_1(i)}^{(i)}} \frac{d J_{i p_1(i)}^{(i)}}{d h_j} + \frac{d\left< x_i x_{p_1(i)}\right>}{d J_{i p_2(i)}^{(i)}} \frac{d J_{i p_2(i)}^{(i)}}{d h_j}\nonumber \\ &+ \frac{d\left< x_i x_{p_1(i)}\right>}{d \left< x_{p_1(i)}\right>} \frac{d \left< x_{p_1(i)}\right>}{d h_j}+ \frac{d\left< x_i x_{p_1(i)}\right>}{d \left< x_{p_1(i)} x_{p_2(i)}\right>} \frac{d \left< x_{p_1(i)} x_{p_2(i)}\right>}{d h_j} \nonumber \\
    &= \frac{e^{-h_i^{(i)}-J_{ip_1(i)}^{(i)}}}{\left( 1+ e^{-h_i^{(i)}-J_{ip_1(i)}^{(i)}}\right)^2}\left(\left< x_{p_1(i)}\right>-\left< x_{p_1(i)}x_{p_2(i)}\right>\right) \left(\frac{d h_i^{(i)}}{d h_j} + \frac{d J_{ip_1(i)}^{(i)}}{d h_j} \right) \nonumber \\
    &+\frac{e^{-h_i^{(i)}-J_{ip_1(i)}^{(i)}-J_{ip_2(i)}^{(i)}}}{\left( 1+ e^{-h_i^{(i)}-J_{ip_1(i)}^{(i)}-J_{ip_2(i)}^{(i)}}\right)^2}\left< x_{p_1(i)}x_{p_2(i)}\right>\left(\frac{d h_i^{(i)}}{d h_j} + \frac{d J_{ip_1(i)}^{(i)}}{d h_j} + \frac{d J_{ip_2(i)}^{(i)}}{d h_j} \right) \nonumber\\
    +& \left[\frac{-1}{ 1+ e^{-h_i^{(i)}}} + \frac{1}{ 1+ e^{-h_i^{(i)}-J_{ip_1(i)}^{(i)}}} \right] \frac{d \left< x_{p_1(i)}\right>}{d h_j} \nonumber \\
    +& \left[\frac{1}{ 1+ e^{-h_i^{(i)}-J_{ip_1(i)}^{(i)}}}  +\frac{1}{ 1+ e^{-h_i^{(i)}-J_{ip_1(i)}^{(i)}-J_{ip_2(i)}^{(i)}}} \right]\frac{d\left< x_{p_1(i)} x_{p_2(i)}\right>}{d h_j}
\end{align}
and
\begin{align}
    \frac{d\left< x_i x_{p_2(i)}\right>}{d h_j} &= \frac{d\left< x_i x_{p_2(i)}\right>}{d h_i ^{(i)}} \frac{d h_i^{(i)}}{ d h_j }+ \frac{d\left< x_i x_{p_2(i)}\right>}{d J_{i p_1(i)}^{(i)}} \frac{d J_{i p_1(i)}^{(i)}}{d h_j} + \frac{d\left< x_i x_{p_2(i)}\right>}{d J_{i p_2(i)}^{(i)}} \frac{d J_{i p_2(i)}^{(i)}}{d h_j}\nonumber \\ 
    &+ \frac{d\left< x_i x_{p_2(i)}\right>}{d \left< x_{p_2(i)}\right>} \frac{d \left< x_{p_2(i)}\right>}{d h_j}+ \frac{d\left< x_i x_{p_2(i)}\right>}{d \left< x_{p_1(i)} x_{p_2(i)}\right>} \frac{d \left< x_{p_1(i)} x_{p_2(i)}\right>}{d h_j} \nonumber \\
    &= \frac{e^{-h_i^{(i)}-J_{ip_2(i)}^{(i)}}}{\left( 1+ e^{-h_i^{(i)}-J_{ip_2(i)}^{(i)}}\right)^2}\left(\left< x_{p_2(i)}\right>-\left< x_{p_1(i)}x_{p_2(i)}\right>\right) \left(\frac{d h_i^{(i)}}{d h_j} + \frac{d J_{ip_2(i)}^{(i)}}{d h_j} \right) \nonumber \\
    &+\frac{e^{-h_i^{(i)}-J_{ip_1(i)}^{(i)}-J_{ip_2(i)}^{(i)}}}{\left( 1+ e^{-h_i^{(i)}-J_{ip_1(i)}^{(i)}-J_{ip_2(i)}^{(i)}}\right)^2}\left< x_{p_1(i)}x_{p_2(i)}\right>\left(\frac{d h_i^{(i)}}{d h_j} + \frac{d J_{ip_1(i)}^{(i)}}{d h_j} + \frac{d J_{ip_2(i)}^{(i)}}{d h_j} \right) \nonumber\\
    +& \left[\frac{-1}{ 1+ e^{-h_i^{(i)}}} + \frac{1}{ 1+ e^{-h_i^{(i)}-J_{ip_2(i)}^{(i)}}} \right] \frac{d \left< x_{p_2(i)}\right>}{d h_j} \nonumber \\
    +& \left[\frac{1}{ 1+ e^{-h_i^{(i)}-J_{ip_2(i)}^{(i)}}}  +\frac{1}{ 1+ e^{-h_i^{(i)}-J_{ip_1(i)}^{(i)}-J_{ip_2(i)}^{(i)}}} \right]\frac{d\left< x_{p_1(i)} x_{p_2(i)}\right>}{d h_j}
\end{align}
Where we directly calculated the derivatives $\langle x_i x_{p_1(i)} \rangle, \langle x_i x_{p_2(i)} \rangle$ with respect to $h_i^{(i)}, J_{i p_1(i)}^{(i)}, J_{i p_2(i)}^{(i)}, \langle x_{p_1(i)}\rangle, \langle x_{p_2(i)}\rangle, \langle x_{p_1(i)} x_{p_2(i)}\rangle$  using Eqs.~[21,22] in the main text. These derivatives are

\begin{equation}
    \frac{d\langle x_i x_{p_1(i)} \rangle}{dh_i^{(i)}} = \frac{e^{-h_i^{(i)}-J_{ip_1(i)}^{(i)}}}{\left( 1+ e^{-h_i^{(i)}-J_{ip_1(i)}^{(i)}}\right)^2}\left(\left< x_{p_1(i)}\right>-\left< x_{p_1(i)}x_{p_2(i)}\right>\right) + \frac{e^{-h_i^{(i)}-J_{ip_1(i)}^{(i)}-J_{ip_2(i)}^{(i)}}}{\left( 1+ e^{-h_i^{(i)}-J_{ip_1(i)}^{(i)}-J_{ip_2(i)}^{(i)}}\right)^2}\left< x_{p_1(i)}x_{p_2(i)}\right>,
\end{equation}
\begin{equation}
    \frac{d\langle x_i x_{p_1(i)} \rangle}{dJ_{i p_1(i)}^{(i)}} = \frac{e^{-h_i^{(i)}-J_{ip_1(i)}^{(i)}}}{\left( 1+ e^{-h_i^{(i)}-J_{ip_1(i)}^{(i)}}\right)^2}\left(\left< x_{p_1(i)}\right>-\left< x_{p_1(i)}x_{p_2(i)}\right>\right) + \frac{e^{-h_i^{(i)}-J_{ip_1(i)}^{(i)}-J_{ip_2(i)}^{(i)}}}{\left( 1+ e^{-h_i^{(i)}-J_{ip_1(i)}^{(i)}-J_{ip_2(i)}^{(i)}}\right)^2}\left< x_{p_1(i)}x_{p_2(i)}\right>,
\end{equation}
\begin{equation}
    \frac{d\langle x_i x_{p_1(i)} \rangle}{dJ_{i p_2(i)}^{(i)}} = \frac{e^{-h_i^{(i)}-J_{ip_1(i)}^{(i)}-J_{ip_2(i)}^{(i)}}}{\left( 1+ e^{-h_i^{(i)}-J_{ip_1(i)}^{(i)}-J_{ip_2(i)}^{(i)}}\right)^2}\left< x_{p_1(i)}x_{p_2(i)}\right>,
\end{equation}
\begin{equation}
    \frac{d\langle x_i x_{p_1(i)} \rangle}{d \langle x_{p_1(i)}\rangle} =\frac{-1}{ 1+ e^{-h_i^{(i)}}} + \frac{1}{ 1+ e^{-h_i^{(i)}-J_{ip_1(i)}^{(i)}}} ,
\end{equation}
\begin{equation}
    \frac{d\langle x_i x_{p_1(i)} \rangle}{d \langle x_{p_2(i)}\rangle} = 0,
\end{equation}
\begin{equation}
    \frac{d\langle x_i x_{p_1(i)} \rangle}{d \langle x_{p_1(i)} x_{p_2(i)}\rangle} =\frac{1}{ 1+ e^{-h_i^{(i)}-J_{ip_1(i)}^{(i)}}}  +\frac{1}{ 1+ e^{-h_i^{(i)}-J_{ip_1(i)}^{(i)}-J_{ip_2(i)}^{(i)}}},
\end{equation}
\begin{equation}
    \frac{d\langle x_i x_{p_2(i)} \rangle}{dh_i^{(i)}} = \frac{e^{-h_i^{(i)}-J_{ip_2(i)}^{(i)}}}{\left( 1+ e^{-h_i^{(i)}-J_{ip_2(i)}^{(i)}}\right)^2}\left(\left< x_{p_2(i)}\right>-\left< x_{p_1(i)}x_{p_2(i)}\right>\right) +\frac{e^{-h_i^{(i)}-J_{ip_1(i)}^{(i)}-J_{ip_2(i)}^{(i)}}}{\left( 1+ e^{-h_i^{(i)}-J_{ip_1(i)}^{(i)}-J_{ip_2(i)}^{(i)}}\right)^2}\left< x_{p_1(i)}x_{p_2(i)}\right>,
\end{equation}
\begin{equation}
    \frac{d\langle x_i x_{p_2(i)} \rangle}{dJ_{i p_1(i)}^{(i)}} = \frac{e^{-h_i^{(i)}-J_{ip_1(i)}^{(i)}-J_{ip_2(i)}^{(i)}}}{\left( 1+ e^{-h_i^{(i)}-J_{ip_1(i)}^{(i)}-J_{ip_2(i)}^{(i)}}\right)^2}\left< x_{p_1(i)}x_{p_2(i)}\right> ,
\end{equation}
\begin{equation}
    \frac{d\langle x_i x_{p_2(i)} \rangle}{dJ_{i p_2(i)}^{(i)}} = \frac{e^{-h_i^{(i)}-J_{ip_2(i)}^{(i)}}}{\left( 1+ e^{-h_i^{(i)}-J_{ip_2(i)}^{(i)}}\right)^2}\left(\left< x_{p_2(i)}\right>-\left< x_{p_1(i)}x_{p_2(i)}\right>\right) +\frac{e^{-h_i^{(i)}-J_{ip_1(i)}^{(i)}-J_{ip_2(i)}^{(i)}}}{\left( 1+ e^{-h_i^{(i)}-J_{ip_1(i)}^{(i)}-J_{ip_2(i)}^{(i)}}\right)^2}\left< x_{p_1(i)}x_{p_2(i)}\right>,
\end{equation}
\begin{equation}
    \frac{d\langle x_i x_{p_2(i)} \rangle}{d \langle x_{p_1(i)}\rangle} = 0,
\end{equation}
\begin{equation}
    \frac{d\langle x_i x_{p_2(i)} \rangle}{d \langle x_{p_2(i)}\rangle} =\frac{-1}{ 1+ e^{-h_i^{(i)}}} + \frac{1}{ 1+ e^{-h_i^{(i)}-J_{ip_2(i)}^{(i)}}} ,
\end{equation}
\begin{equation}
    \frac{d\langle x_i x_{p_2(i)} \rangle}{d \langle x_{p_1(i)} x_{p_2(i)}\rangle} = \frac{1}{ 1+ e^{-h_i^{(i)}-J_{ip_2(i)}^{(i)}}}  +\frac{1}{ 1+ e^{-h_i^{(i)}-J_{ip_1(i)}^{(i)}-J_{ip_2(i)}^{(i)}}} .
\end{equation}
This ends the calculation for the derivatives of $\left< x_i\right> , \left< x_i x_{p_1(i)}\right>, \left< x_i x_{p_2(i)}\right>$ with respect to $h_j$

We can now repeat this calculation for the covariance between node $j$ and another node $k>j$ to find the correlation $\langle x_k x_j\rangle$. Once we calculate all possible correlations $\langle x_k x_j\rangle$ for $k = N ... j+1$ we can again repeat this calculation for node $j-1$ and all previously added nodes. We continue until we compute all pairwise correlations $\langle x_i x_j \rangle$ between variables.

\subsection*{Frustration Synergy Bound}

Synergy represents the amount of information that two variables contain about a third above and beyond their pairwise mutual information. Synergy can be written as
\begin{equation}
    \text{Syn}(x_i,x_j,x_k) = S(x_i)+S(x_j)+S(x_k) - S(x_i,x_j,x_k)-I(x_i,x_j)-I(x_i,x_k)-I(x_j,x_k).
    \label{EQ:Synergy}
\end{equation}
When the variables are modeled by a Ising model
\begin{equation}
    P_G = \frac{1}{Z} \exp \left(  J_{ij} x_i x_j+J_{ik} x_i x_k+J_{jk} x_j x_k +  h_i x_i+h_j x_j+h_k x_k \right), 
\end{equation}
each term in synergy depends on the Ising interaction $J_{ij},J_{ik},J_{jk}$ between variables and local fields $h_i,h_j,h_k$.  If the product of interactions is negative the system will be \textit{frustrated}. Formally, we defined frustration as $F = - J_{ij} J_{ik} J_{jk}$, where $F$ positive is a system that is frustrated. Predicting the activity of one of the variables from the other two will be worse in a frustrated system than in a non-frustrated system. Therefore, we should expect the synergy to be higher for a frustrated system than a non-frustrated system as the information contained in the full model must be higher than the pairwise mutual information between the three variables. We investigate how the synergy depends on the level of frustration within the model. Specifically, given a system that has some level of frustration, what is the maximal achievable synergy for this system? We employ the method of Lagrange multipliers to maximize synergy subject to a fixed value of frustration, maximizing the Lagrangian
\begin{equation}
    \mathcal{L} = \text{Syn} (J_{12},J_{13},J_{23}, h_1, h_2, h_3) - \lambda (F - J_{12}J_{13}J_{23}),
\end{equation}
where $\text{Syn} (J_{12},J_{13},J_{23}, h_1, h_2, h_3) $ is a function of Ising model parameters and $F$ is the frustration we are constraining the system. We perform this analysis numerically for each value of frustration. This is how we constructed the bound for the red forbidden region in Fig.~5C in the main text. We can see by this bound that positive synergy is only possible when the system is frustrated. 

\subsection*{Overlap Between Random Networks} 

\begin{figure}
    \centering
    \includegraphics[width=\textwidth]{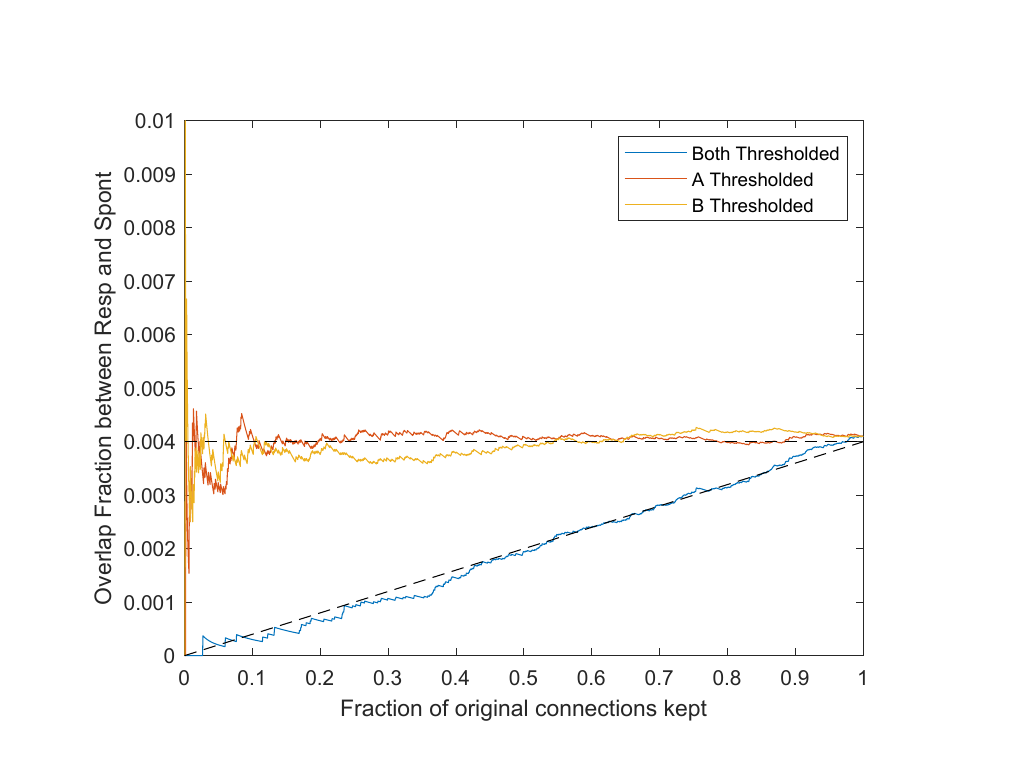}
    \caption{Fractional Overlap Simulation vs. Theory. 50 random models of 1000 neurons were simulated. The plot above shows the average over all models.}
    \label{fig:fracoverlap}
\end{figure}

In the main text, we have data from neurons responding from different stimuli. Given that we have pairs of the same neurons responding to different stimuli, we want to compare how consistent the optimal correlations are between the two stimuli. To do this, we need to compute the overlap between the two networks. Consider two networks with $N$ nodes each, $E_1$ and $E_2$ edges, and adjacency matrices $G^{(1)}$ and $G^{(2)}$. We want to compute the number of overlapping edges,
\begin{equation}
E_{12} = \sum_{(ij)} G^{(1)}_{ij}G^{(2)}_{ij}.
\end{equation}
If the two networks are independent, then the expected fraction of edges in $G^{(2)}$ that are also in $G^{(1)}$ is given by
\begin{equation}
\frac{\bar{E_{12}}}{E_2} = \frac{2E_1}{N(N-1)}.
\end{equation}
The expected number of shared edges is therefore
\begin{equation}
\bar{E_{12}} = \frac{2E_1E_2}{N(N-1)}.
\end{equation}
We confirm these predictions in random networks (Fig.~\ref{fig:fracoverlap}).

\subsection*{Topological Structure of Networks}

\begin{figure}
    \centering
    \includegraphics[width=\textwidth]{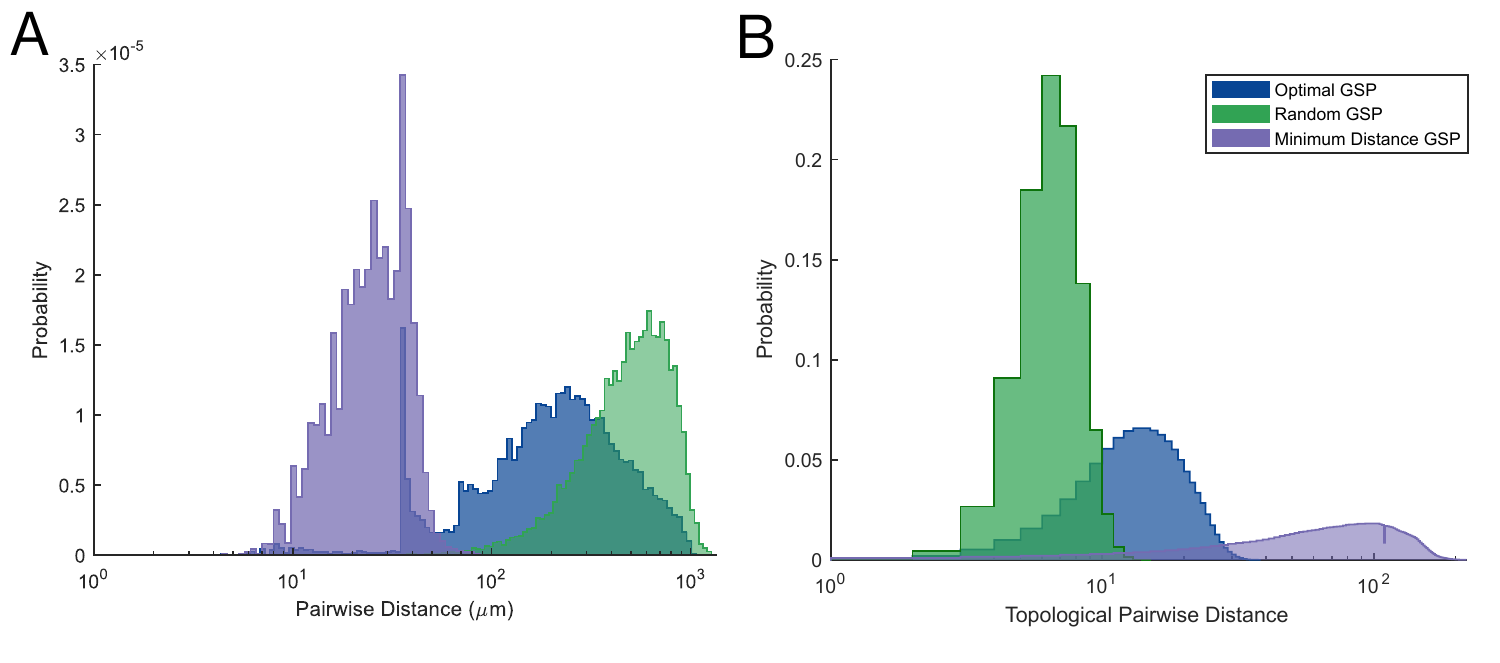}
    \caption{Topological and Physical Distance distribution. (A) The Physical distances between connected neurons. (B) Topological Distance (number of hops) between neurons.}
    \label{fig:Topo}
\end{figure}

In addition to studying the predictions of the minimax entropy model $P_G$, we can examine the topological and physical structures of optimal correlation networks $G$. The brain must balance maximizing neuronal communication and minimizing energy expenditure for these pathways \cite{harris_synaptic_2012}. Using a population of 10,144 neurons recorded while the mouse is exposed to a sequence of natural images, the same example population used in the main text, we investigate where the optimal network lies between communication and energy efficiency. Given the physical location of each neuron, we can calculate inter-neuron distances in the correlation network. Networks with short physical connections typically require more hops to traverse, while those minimizing topological distance (the small-world effect \cite{bullmore_complex_2009,bullmore_economy_2012,sporns_contributions_2014}) necessitate longer physical connections. We compare the optimal network to two alternative network types: a minimum distance GSP network that minimizes neuron connection distances, and a random GSP network that minimizes topological steps. Figs. \ref{fig:Topo}A and B compare the optimal network's physical and topological distance distributions with these two network types, each containing the same number of edges as the optimal network. The minimax entropy model identifies connections that are shorter than average (Fig. \ref{fig:Topo}A) when compared to the random network. In Fig. \ref{fig:Topo}B minimax entropy model finds a network that has shorter topological distances than the equivalent-size minimum distance network. Therefore, the minimax entropy network identifies short physical connections that maintain the small-world structure observed in real neuronal networks.

\subsection*{Scaling of Information}

\begin{figure}
    \centering
    \includegraphics[width=\textwidth]{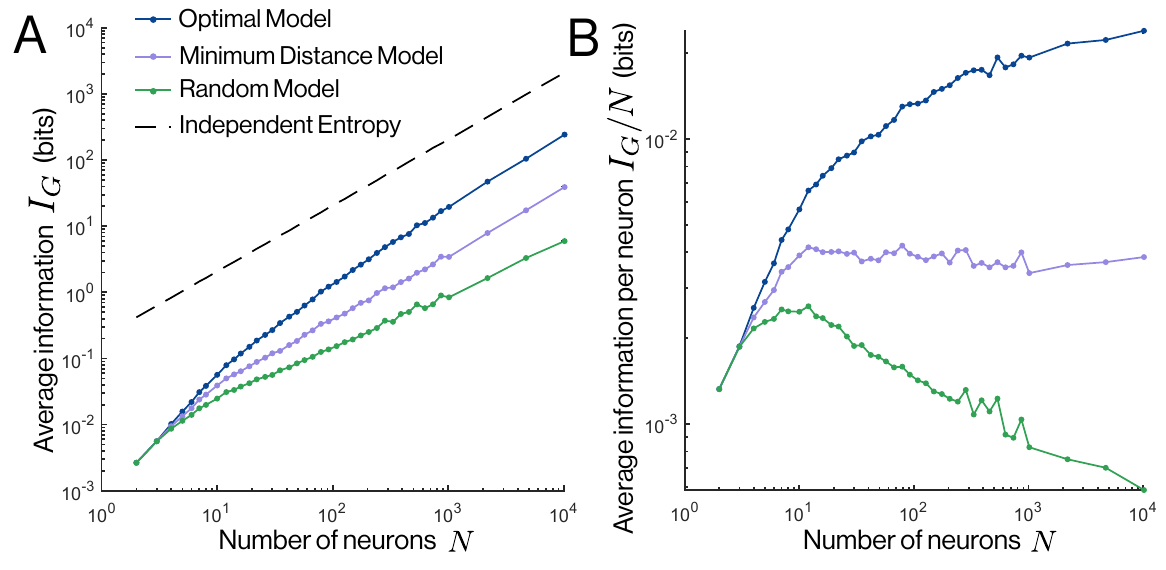}
    \caption{Scaling of optimal model. (A) The information contained in the model is the difference between the model entropy and the independent entropy of the data. Shown is the information contained in the network of correlations for the minimax entropy model, a model constructed by minimizing the distance between neurons, and random models. (B) The fractional information of the model as a function of neuron population size. As the population grows, the fractional information continues to increase for the optimal network, while the minimum distance network becomes static and the random network decreases.}
    \label{fig:scaling}
\end{figure}

In the main text, we analyze our model's performance on a single experimental recording of approximately 10,000 neurons. As experiments probe even larger populations, it is unclear whether the minimax entropy model will scale effectively with the growing population size. To understand how our model scales, we implement a subsampling approach. We begin by making subpopulations of size $N$ from our total population of neurons. We construct our subpopulations by starting with a central neuron and adding neurons to the subpopulation that are closest to our starting neuron. Then for each subpopulation of size $N$, we pick starting neurons so that the aggregate subpopulations of this size span the entire population. We then applied our minimax entropy method to these subpopulations and calculated the information in the optimal network for these subpopulations. Finally, for each subpopulation size $N$, we average the results over all subpopulations. As the number of neurons increases, the independent entropy increases linearly (Fig.~\ref{fig:scaling}A). Each GSP network contains $2N -3$ correlations. Thus, one might also expect the information contained in each network to grow linearly with $N$. We see from this figure that information in all three networks grows with $N$ but the rate at which they grow diverges for systems larger than 10 neurons. 

If the information $I_G$ grows linearly with the population size $N$, then the information per neuron explained by the correlations in $G$, $I_G/N$, must be constant in $N$. For the minimum distance GSP network, we find that the information per neuron plateaus at approximately 0.004 bits for populations larger than 10 neurons (Fig.~\ref{fig:scaling}B). In contrast, a random GSP network exhibits a decreasing information per neuron. To understand this difference, we need to examine Fig.~2A in the main text and consider how our models choose their respective networks. The distribution of entropy drops is heavy-tailed. Therefore, the random network, which represents an average GSP network that can be constructed from our data, will predominantly select uninformative connections. A minimum distance network will fare better as neurons close to each other contain information about the activity of their neighbors. This is why the information per neuron increases for small populations for both the minimum distance network and the random network but eventually loses out to the informative long-range correlations that these models ignore. As the population size grows the minimax model chooses connections from the tail of the entropy drop distribution. Therefore, the information gain of the minimax model will grow superlinear with the population size. The optimal model in Fig.~\ref{fig:scaling} B does not plateau at the full population size suggesting that our model would continue to perform well for even larger populations. 

\subsection*{Comparison to Random Networks}

\begin{figure}
    \centering
    \includegraphics[width=\textwidth]{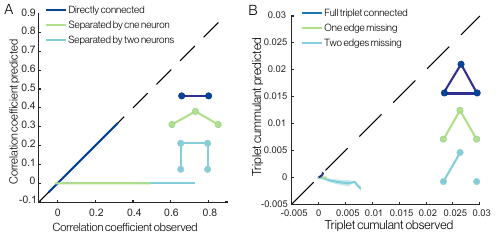}
    \caption{Accuracy of random. (A) For the same recording of N = 10, 144 neurons as Fig. 4 in the main text, we compare the correlation coefficients between pairs of neurons measured in
data versus predicted by the random maximum entropy model. Different colors represent pairs of neurons separated by different distances in the network G, and the dashed line
indicates equality. Plots are computed by binning neuron pairs along the x-axis, with lines and shaded regions representing means and standard deviations within each bin. (B)
Cumulants among triplets of neurons measured in data versus predicted by a random model. Different colors represent triplets of neurons with different numbers of
pairwise correlations constrained in the model. Lines and shaded regions represent means and standard deviations within bins along the x-axis}
    \label{fig:random-model-accuracy}
\end{figure}

In order to test the performance of our optimal network, we compare it to a random network. As shown in Fig. \ref{fig:random-model-accuracy}A, a random network cannot predict the correlations that are not included in the network. For higher-order statistics, Fig. \ref{fig:random-model-accuracy}B demonstrates that the random network provides negligible predictive power. Comparing these results with Fig. 4 in the main text demonstrates that our optimal network chooses statistics that are predictive of statistics that are not part of our network. 

\begin{figure}
    \centering
    \includegraphics[width=\textwidth]{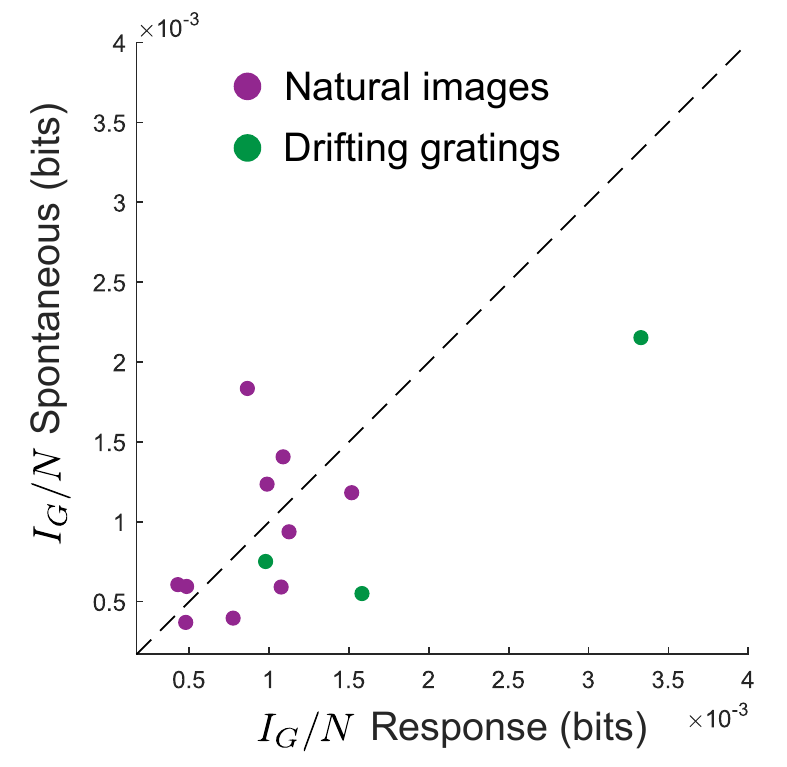}
    \caption{Information between random networks during visual responses and
spontaneous activity. The same network was used for each pair of responding and spontaneous activity.}
    \label{fig:spontresprandom}
\end{figure}

We also investigated how external stimuli influence the information captured by the optimal network of correlations. In the main text, we compared the information $I_G$ captured by the optimal networks during exposure to visual stimuli versus spontaneous activity. During visual stimulation, we found that the optimal network of correlations captures even more information $I_G$ than during spontaneous activity. Here, we investigate the same question for random networks of correlations. Fig.~\ref{fig:spontresprandom} shows that random networks capture roughly equal information between visual stimulation and spontaneous activity.

\subsection*{Simulated Data Comparison}

\begin{figure}
    \centering
    \includegraphics[width=\textwidth]{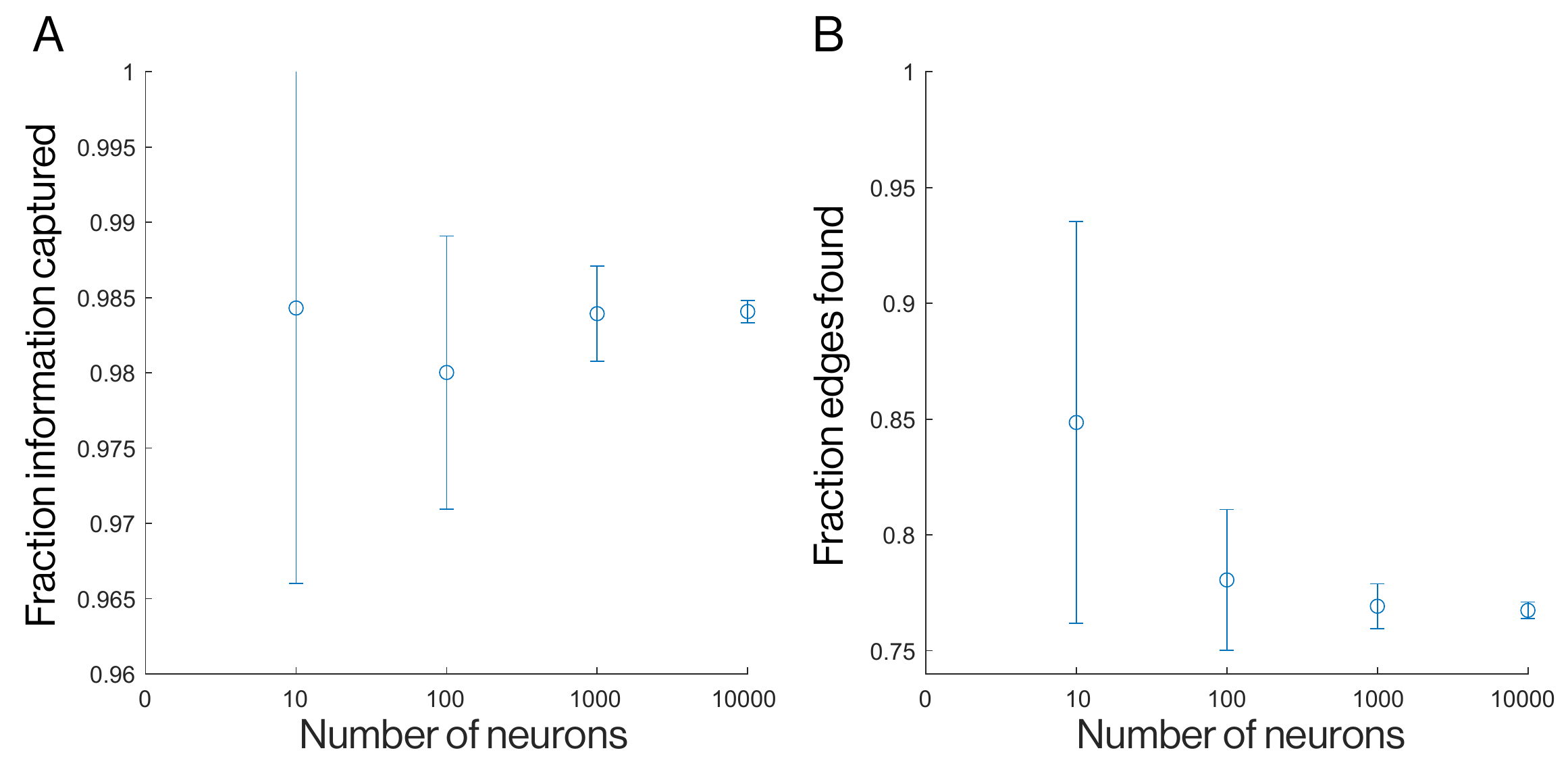}
    \caption{Performance of minimax entropy method compared to simulated data. Simulated networks were generated by creating random GSP networks with Gaussian distributed interaction strengths $J_{ij }$ and external fields $h_i$.  (A) Fraction of information captured by the minimax entropy method from the total information available.  (B) Overlap fraction of edges between the simulated model and the inferred minimax entropy model.}
    \label{fig:gspsimulated}
\end{figure}

We tested the performance of our minimax entropy method by generating random Ising GSP networks with Gaussian-distributed interactions and external fields, $J_{ij}, h_i \sim \mathcal{N}(0,1) $. We varied network sizes from $N = 10$ to $10000$. The number of simulation iterations we averaged over scaled inversely to network size: $100000/N$ iterations. As shown in Fig.~\ref{fig:gspsimulated} we capture 98\% of the information in these simulated models and 75\% of the ground truth interactions, $J_{ij}$.

\begin{table}
\caption{Response and spontaneous table: Listing the 13 experiments where both the responding data and the spontaneous data were compared.}
\centering
\begin{tabular}{llll}
\begin{tabular}[c]{@{}l@{}}Experiment Name\end{tabular} & \begin{tabular}[c]{@{}l@{}}Data Response Length\end{tabular} & \begin{tabular}[c]{@{}l@{}}Data Spontaneous Length\end{tabular} & \begin{tabular}[c]{@{}l@{}}Number of Neurons\end{tabular}  \\
\midrule
natimg2800\_M160825\_MP027\_2016-12-14                    & 5426                                                          & 4696                                                             & 11449                                                       \\
natimg2800\_M161025\_MP030\_2017-05-29                    & 5658                                                          & 2565                                                             & 14062                                                       \\
natimg2800\_M170714\_MP032\_2017-08-07                    & 5880                                                          & 2586                                                             & 9039                                                        \\
natimg2800\_8D\_M170604\_MP031\_2017-07-02                & 5880                                                          & 2763                                                             & 9147                                                        \\
natimg2800\_8D\_M170717\_MP033\_2017-08-22                & 5880                                                          & 2584                                                             & 10131                                                       \\
natimg2800\_small\_M170717\_MP033\_2017-08-23             & 5880                                                          & 2585                                                             & 11629                                                       \\
ori32\_M170604\_MP031\_2017-06-26                         & 3300                                                          & 3700                                                             & 8930                                                        \\
ori32\_M170714\_MP032\_2017-08-02                         & 1980                                                          & 2229                                                             & 9208                                                        \\
ori32\_M170717\_MP033\_2017-08-17                         & 3300                                                          & 3700                                                             & 8991                                                        \\
natimg32\_M150824\_MP019\_2016-03-23                      & 3000                                                          & 4823                                                             & 11652                                                       \\
natimg32\_M170604\_MP031\_2017-06-27                      & 3300                                                          & 3700                                                             & 11992                                                       \\
natimg32\_M170714\_MP032\_2017-08-01                      & 3762                                                          & 4218                                                             & 12578                                                       \\
natimg32\_M170717\_MP033\_2017-08-25                      & 3168                                                          & 3552                                                             & 11523                                                      
\end{tabular}
\end{table}


\FloatBarrier

\bibliography{ExactPNAS_cwl1}